\documentclass[amssymb,11pt]{article}
\usepackage{amsmath}
\usepackage{amssymb}
\usepackage{graphicx}
\usepackage[ruled]{algorithm2e}

\SetAlFnt{\small}
\SetAlCapFnt{\small}
\SetAlCapNameFnt{\small}
\SetAlCapHSkip{0pt}
\IncMargin{-\parindent}

\usepackage{times}
\usepackage{amsmath}
\usepackage{stmaryrd}
\usepackage{amssymb}
\usepackage{amssymb}
\usepackage{graphicx}

\newtheorem{thm}{Theorem}[section]

\newtheorem{lem}{Lemma}[section]
\newtheorem{rem}{Remark}[section]
\newtheorem{defn}{Definition}[section]
\newtheorem{prop}{Proposition}[section]

\newtheorem{exam}{Example}[section]

\def\>{\ensuremath{\rangle}}
\def\<{\ensuremath{\langle}}

\begin{document}

\title{Alternation in Quantum Programming: \\ From Superposition of Data to Superposition of Programs}

\author{Mingsheng Ying, Nengkun Yu and Yuan Feng\thanks{Mingsheng Ying and Yuan Feng are with University of Technology, Sydney and Tsinghua University; Nengkun Yu is with University of Waterloo. This work was partly supported by the Australian Research Council (Grant No: DP110103473 and DP130102764).
\small \em Email: Mingsheng.Ying@uts.edu.au or yingmsh@tsinghua.edu.cn}}
\date{}
\maketitle

\begin{abstract} We extract a novel quantum programming paradigm - superposition of programs - from the design idea of a popular class of quantum algorithms, namely quantum walk-based algorithms. The generality of this paradigm is guaranteed by the universality of quantum walks as a computational model.
A new quantum programming language QGCL is then proposed to support the paradigm of superposition of programs. This language can be seen as a quantum extension of Dijkstra's GCL (Guarded Command Language). Surprisingly, alternation in GCL splits into two different notions in the quantum setting: classical alternation (of quantum programs) and quantum alternation, with the latter being introduced in QGCL for the first time. Quantum alternation is the key program construct for realizing the paradigm of superposition of programs.

The denotational semantics of QGCL are defined by introducing a new mathematical tool called the guarded composition of operator-valued functions. Then the weakest precondition semantics of QGCL can straightforwardly derived. Another very useful program construct in realizing the quantum programming paradigm of superposition of programs, called quantum choice, can be easily defined in terms of quantum alternation. The relation between quantum choices and probabilistic choices is clarified through defining the notion of local variables. We derive a family of algebraic laws for QGCL programs that can be used in program verification, transformations and compilation. The expressive power of QGCL is illustrated by several examples where various variants and generalizations of quantum walks are conveniently expressed using quantum alternation and quantum choice. We believe that quantum programming with quantum alternation and choice will play an important role in further exploiting the power of quantum computing.
\end{abstract}

\smallskip\


\bigskip\

\textit{\textquotedblleft I suggested the notation for
a \textquoteleft case expression\textquoteright which selects between any number of alternatives according to the value of an integer
expression. That was my second language design proposal.
I am still most proud of it, because it raises
essentially no problems either for the implementor, the
programmer, or the reader of a program.\textquotedblright}

-- A. C. R. Hoare, The emperor's old clothes, \textit{Communications of the ACM} 24(1981)75-83.

\smallskip\

\section{Introduction}\label{Intro}

Since Knill~\cite{Kn96} introduced the Quantum Random Access Machine (QRAM) model for quantum computing and proposed a set of conventions for writing quantum pseudo-codes in 1996, several  high-level quantum programming languages have been defined in the last 17 years; for example imperative languages QCL by \"{O}mer~\cite{O03} and qGCL by Sanders and Zuliani~\cite{SZ00,Z01}, and functional languages QPL by Selinger~\cite{Se04} and QML by Altenkirch and Grattage~\cite{AG05}. Also, Tafliovich and Hehner~\cite{TH06} defined a quantum extension of Hehner's predicative programming language. Various semantics have been introduced for quantum programming languages; for example, D'Hondt and Panangaden \cite{DP06} introduced the notion of quantum weakest precondition, and a quantum predicate transformer semantics was proposed in \cite{YD10}. Several proof systems for verification of quantum programs have been developed; for example Baltag and Smets \cite{BS06} presented a dynamic logic of quantum information flow, Brunet and Jorrand \cite{BJ04} introduced a way of using Birkhoff-von Neumann quantum logic to reason about quantum programs, and Hoare logic was generalized to prove both partial and total correctness of quantum programs by Chadha, Mateus and Sernadas~\cite{Ch06}, Kakutani \cite{Ka09} and the authors~\cite{FY07,Y11}. The implementation of quantum programming languages has also attracted attention \cite{Sv06,Na07,YF11} as the rapid progress of quantum technology has made people widely believe that large-scalable and functional quantum computers will be built in not too far future.
An excellent survey of research on quantum programming before 2006 can be found in~\cite{Ga06}, and for a more recent survey, see~\cite{Ying10,Ying12}. It is particularly worth pointing out that three more practical quantum programming languages were announced in the last two years: two general-purpose languages Quipper by Green, LeFanu Lumsdaine, Ross, Selinger and Valiron~\cite{Gre13}, and Scaffold by Abhari, Faruque, Dousti, et. al.~\cite{Abh12}, and a domain-specific language QuaFL by Lapets, da Silva, Thome, Adler, Beal and R\"{o}tteler~\cite{Lap13}.

Now the development of the theory of quantum programming has reached such a stage that the quantum extensions of various basic program constructs (e.g. sequential composition) have been properly introduced in the languages mentioned above. Then an important problem for further studies would be to (re)examine more sophisticated program constructs and programming abstractions and models that have been successfully used in classical programming in the quantum setting: how to define the quantum counterparts of them? how can they be used in programming a quantum computer? is it possible to employ them to solve a problem more efficiently on a quantum computer than on a classical computer? A further problem is: what kind of new programming language features that have not been introduced or even irrelevant in classical programming are needed in order to exploit the full capability of a quantum computer?

Alternation, case statement or switch statement is a very convenient program construct to implement a case analysis in classical programming~\cite{Di75,Mor98}. Thus, most high-level imperative programming languages possess alternation constructs. In particular, the nondeterminism derived from alternation provides a basis for refinement-based program development; for example, Barman, Bod\'ik, Chandra, Galenson, Kimelman, Rodarmor and Tung~\cite{Bar10} recently introduced a methodology of programming with angelic nondeterminism. This paper identifies a novel quantum programming paradigm - superposition of programs - from the design idea of a class of popular quantum algorithms, namely quantum walk-based algorithms. We find that a quantum generalization of alternation is crucial to support quantum programming in this new paradigm. Surprisingly, the notion of alternation in classical programming languages splits into two different notions in the quantum setting: classical alternation (of quantum programs) and quantum alternation. Classical alternation of quantum programs has already been properly introduced in the previous works on quantum programming, but it is not the program construct that we require for the purpose of realizing superposition of programs. The major aim of this paper is to define the new notion of quantum alternation that can support the paradigm of superposition of programs.

\subsection{Alternation and Choice in Classical Programming}\label{12}
 Recall that an alternation is a collection of guarded commands written as \begin{equation}\label{altn}\begin{split}\mathbf{if}\ &G_1\rightarrow P_1\\ \square\ &G_2\rightarrow P_2\\ &\ \ \ ...... \\ \square\ &G_n\rightarrow P_n\\ \mathbf{fi}\ &
\end{split}\end{equation} or \begin{equation}\label{altn1}\mathbf{if}\ (\square i\cdot G_i\rightarrow P_i)\ \mathbf{fi}\end{equation} where for each $1\leq i\leq n$, the subprogram $P_i$ is guarded by the boolean expression $G_i$, and $P_i$ will be executed only when $G_i$ is true.

Alternation is also the most widely accepted mechanism for nondeterministic programming. Nondeterminism in alternation~(\ref{altn}) or (\ref{altn1}) is a consequence of the \textquotedblleft overlapping" of the guards $G_1,G_2,...,G_n$; that is, if more than one guards $G_i$ are true at the same time, the alternation needs to select one from the corresponding commands $P_i$ for execution. In particular, if $G_1=G_2=\cdots=G_n=\mathbf{true}$, then alternation~(\ref{altn}) or (\ref{altn1}) becomes a demonic choice: \begin{equation}\label{dech}\square_{i=1}^n\ P_i\end{equation} where the alternatives $P_i$ are chosen unpredictably.

To formalise randomised algorithms, research on probabilistic programming \cite{Ko81,MM05} started in 1980's with the introduction of probabilistic choice:\begin{equation}\label{proch}\square_{i=1}^n\ C_i@p_i\end{equation} where $\{p_i\}$ is a probability distribution; that is, $p_i\geq 0$ for all $i$, and $\sum_{i=1}^n p_i=1$. The probabilistic choice~(\ref{proch}) randomly chooses the command $C_i$ with probability $p_i$ for every $i$, and thus it can be seen as a refinement of the demonic choice (\ref{dech}). A probabilistic choice is often used to represent a decision in forks according to a certain probability distribution in a randomised algorithm.

\subsection{Classical Alternation in Quantum Programming}
 As stated before, the aim of this paper is to define a quantum generalization of alternation. Indeed, a kind of alternation already exists in Sanders and Zuliani's quantum programming language qGCL~\cite{SZ00,Z01} because qGCL is the probabilistic GCL \cite{MM05} extended by adding the quantum procedures of unitary transformations and measurements, and thus alternation and probabilistic choice in pGCL are inherited in qGCL. Another kind of measurement-based alternation was introduced by Selinger in his quantum programming language QPL~\cite{Se04}. Let $\overline{q}$ be a family of quantum variables and $M$ a measurement on $\overline{q}$ with possible outcomes $m_1,m_2,...,m_n$. For each $1\leq i\leq n$, let $P_i$ be a (quantum) program. Then a generalized form of Selinger's alternation considered in \cite{Y11} can be written as follows:

\begin{equation}\label{maltn}\begin{split}&\mathbf{measure}\ M[\overline{q}]=\ m_1\rightarrow P_1\\ &\ \ \ \ \ \ \ \ \square\ \ \ \ \ \ \ \ \ \ \ \ \ \ \ \ \ \ \ \ \ m_2\rightarrow P_2\\ &\ \ \ \ \ \ \ \ \ \ \ \ \ \ \ \ \ \ \ \ \ \ \ \ \ \ \ \ \ \ \ \ \ \ \ \ \ ...... \\ &\ \ \ \ \ \ \ \ \square\ \ \ \ \ \ \ \ \ \ \ \ \ \ \ \ \ \ \ \ \ m_n\rightarrow P_n\\ &\mathbf{end}
\end{split}\end{equation} or \begin{equation}\label{maltn1}\mathbf{measure}\ (\square i\cdot M[\overline{q}]=m_i\rightarrow P_i)\ \mathbf{end}\end{equation}
Alternation (\ref{maltn}) or (\ref{maltn1}) selects a command according to the outcome of measurement $M$: if the outcome is $m_i$, then the corresponding command $P_i$ will be executed.
The alternations defined in both qGCL and QPL can be appropriately termed as \textit{classical alternation of quantum programs} because the selection of commands in it based on classical information - the outcomes of quantum measurements. However, our intention is to introduce the notion of \textit{quantum alternation of (quantum) programs}. Do we actually need quantum alternation in quantum programming? A role for programming languages is to provide ways of organizing computations~\cite{Se02}. So, to answer this question, let's look at the basic design ideas of several popular quantum algorithms.

\subsection{From Superposition of Data to Superposition of Programs}

\subsubsection{Superposition of Data}
It has been realized well that one major source of the power of quantum computation \cite{NC00} is superposition of data. To see this, let's consider a function $$f(x_1,...,x_n):\{0,1\}^n\rightarrow\{0,1\}$$ with $n-$bit input and one-bit output, we want to compute $f(x)$ for multiple inputs $x=(x_1,...,x_n)\in\{0,1\}^n$ simultaneously. Classical parallelism is meant to build \textit{multiple} circuits all for computing the same function $f$ and to execute them in parallel for different inputs. However, quantum parallelism allows us to computer $f(x)$ for all different inputs $x\in\{0,1\}^n$ simultaneously with a \textit{single} quantum circuit implementing the oracle unitary operator: \begin{equation}\label{OR}U_f:|x,y\rangle\rightarrow |x,y\oplus f(x)\rangle\end{equation} where $x\in\{0,1\}^n$, $y\in\{0,1\}$ and $\oplus$ is addition module $2$. This quantum circuit has $n+1$ qubits: the $n$ qubits $x=(x_1,...,x_n)$ are called the data register, and the qubit $y$ is called the target register. Initially, we prepare the data and target registers in computational basis state $|0\rangle^{\otimes n}|0\rangle$. The superposition \begin{equation}\label{dsup}|\psi\rangle\stackrel{\triangle}{=}\frac{1}{\sqrt{2^n}}\sum_{x\in\{0,1\}^n}|x\rangle=H^{\otimes n}|0\rangle^{\otimes n}\end{equation} of the basis states of $n$ qubits can be created by $n$ Hadamard gates $H^{\otimes n}$ acting in parallel on the data register, where \begin{equation}\label{HDG}H=\frac{1}{\sqrt{2}}\left(\begin{array}{cc}1 & 1\\ 1&-1\end{array}\right)\end{equation} is the Hadamard gate. Then we apply unitary operator $U_f$: $$|\psi\rangle|0\rangle\stackrel{U_f}{\longrightarrow}\frac{1}{\sqrt{2^n}}\sum_{x\in\{0,1\}^n}|x\rangle|f(x)\rangle\stackrel{\triangle}{=}|\varphi\rangle.$$ The different terms of $|\varphi\rangle$ contain information about values $f(x)$ for all $2^n$ inputs $x\in\{0,1\}^n$. Thus, we have computed $f(x)$ for all $x\in\{0,1\}^n$ simultaneously by employing $U_f$ only once. Such a technique of applying a single circuit $U_f$ to the superposition (\ref{dsup}) of data is a key step of a large class of quantum algorithms, including
the Grover search algorithm \cite{Gro96}, the Deustch-Josza algorithm \cite{DJ92} and the Bernstein-Vazirani algorithm \cite{BV93}.

\subsubsection{Superposition of Programs}
After understanding superposition of data, a question naturally arises: is there any other form of superposition that is useful in quantum computing?
A superposition of evolutions (rather than that of states) of a quantum systems was considered by physicists Aharonov, Anandan, Popescu  and Vaidman~\cite{AAPV90} as early as in 1990, and they proposed to introduce an external system in order to implement the superposition. The idea of using such an external system was rediscovered by Aharonov, Ambainis, Bach, Kempe, Nayak, Vazirani, Vishwanath and Watrous
in defining quantum walks~\cite{Am01,Ah01}. Let's consider a simple example of the quantum walk on a graph:

\begin{exam}\label{qwk} A quantum walk is the quantum counterpart of a random walk. Let $G=(V,E)$ be an $n-$regular directed graph; that is, a graph where each vertex has $n$ neighbors. Then we can label each edge with a number between $1$ and $n$ such that for each $1\leq i\leq n$, the directed edges labeled $i$ form a permutation. A random walk on $G$ is defined as follows: the vertices $v$'s of $G$ are used to represent the states of the walk, and for each state $v$ the walk goes from $v$ to its every neighbor with a certain probability. To define a quantum walk on $G$, let $\mathcal{H}_V$ be the Hilbert space spanned by states $\{|v\rangle\}_{v\in V}$ corresponding to the vertices of the graph. Then for each $1\leq i\leq n$, we can define a shift operator $S_i$ on $\mathcal{H}_V$: $$S_i|v\rangle=\left |{\rm the}\ i{\rm th\ neighbour\ of}\ v\right\rangle$$ for any $v\in V$. We introduce an auxiliary quantum system with the state Hilbert space $\mathcal{H}_C$ spanned by $\{|i\rangle\}_{i=1}^n$. This auxiliary system is usually called a \textquotedblleft quantum coin\textquotedblright, and the space $\mathcal{H}_C$ is referred to as the \textquotedblleft coin space\textquotedblright. Now we are able to combine these unitary operators $S_i$ $(1\leq i\leq n)$ along the \textquotedblleft coin\textquotedblright\ to form a whole shift operator $S$ on $\mathcal{H}_C\otimes\mathcal{H}_V$: \begin{equation}\label{sht}S|i, v\rangle= |i\rangle S_i|v\rangle\end{equation} for any $1\leq i\leq n$ and $v\in V$.
If we further choose a unitary operator $C$ on $\mathcal{H}_C$, e.g. the Hadamard gate defined by equation (\ref{HDG}), called the \textquotedblleft coin-tossing operator\textquotedblright, then a single step of a coined quantum walk on graph $G$ can be modelled by the unitary operator: \begin{equation}\label{wed}W\stackrel{\triangle}{=}S(C\otimes I_{\mathcal{H}_V})\end{equation} where $I_{\mathcal{H}_V}$ is the identity operator on $\mathcal{H}_V$. The quantum walk is then an iteration of the single-step walk operator $W$.
\end{exam}

Let's carefully observe the behavior of the quantum walk step $W$. The shift operators $S_1,S_2,...,S_n$ can be seen as a collection of programs independent to each other. Then the whole shift operator $S$ can be seen as a kind of alternation of $S_1,S_2,...,S_n$ because $S$ selects one of them for execution. But the defining equation (\ref{sht}) of $S$ clearly indicates that this alternation is different from the alternation (\ref{maltn}) or (\ref{maltn1}): the selection in equation (\ref{sht}) is made according to the basis state $|i\rangle$ of the \textquotedblleft coin space\textquotedblright, which is quantum information rather than classical information. Thus, we can appropriately call $S$ an \textit{quantum alternation}. Furthermore, the \textquotedblleft coin-tossing operator\textquotedblright\ $C$ can be seen as another program. From equation (\ref{wed}) we see that the quantum walk step $W$ first runs \textquotedblleft coin-tossing\textquotedblright\ program $C$ to create a superposition of the execution paths of programs $S_1,S_2,...,S_n$, and then the quantum alternation $S$ follows. During the execution of alternation $S$, each $S_i$ is running along its own path within the whole superposition of execution paths of $S_1,S_2,...,S_n$. Then the quantum walk step $W$ is indeed a \textit{quantum choice} of shift programs $S_1,S_2,...,S_n$ through the \textquotedblleft coin-tossing\textquotedblright\ program $C$. Therefore, the superposition in $W$ is a higher-level superposition - the \textit{superposition of programs} $S_1,S_2,...,S_n$.
\subsection{Design Decision of the Paper}\label{decin} Quantum walks have been shown to be a very powerful tool for the development of a large class of quantum algorithms, in particular for simulation of quantum systems (see~\cite{Ven12} for a comprehensive review). Moreover, they were proved to be a universal model of computation~\cite{Chi09,Lo10}. This motivates us to develop the idea of superposition of programs embedded in quantum walk-based algorithms as a \textit{quantum programming paradigm}. As suggested by Example~\ref{qwk}, the following two steps are needed toward a general form of superposition of programs.
\subsubsection{Quantum Alternation} The key step is to define quantum alternation. This is exactly the major aim of this paper. The defining equation (\ref{sht}) of the shift operator $S$ of a quantum walk already provides us with a basic idea for defining quantum alternation. Let $P_1,P_2,...,P_n$ be a collection of (quantum) programs whose state spaces are the same Hilbert space $\mathcal{H}$. We introduce a new family of quantum variables $\overline{q}$ that do not appear in $P_1,P_2,...,P_n$. These variables are used to denote an external \textquotedblleft coin\textquotedblright\ system. Assume that the state space of system $\overline{q}$ is an $n-$dimensional Hilbert space $\mathcal{H}_C$ and $\{|i\rangle\}_{i=1}^n$ is an orthonormal basis of it. Then it seems that a quantum alternation $P$ of programs $P_1,P_2,...,P_n$ can be defined by combining them along the basis $\{|i\rangle\}$, simply mimicking the shift operator $S$. More precisely, the semantic operator $\llbracket P\rrbracket$ of alternation $P$ should be defined on the tensor product $\mathcal{H}_C\otimes\mathcal{H}$, and
\begin{equation}\label{aldef}\llbracket P\rrbracket (|i\rangle |\varphi\rangle)=|i\rangle (\llbracket P_i\rrbracket|\varphi\rangle)\end{equation} for every $1\leq i\leq n$ and $|\varphi\rangle\in\mathcal{H}$, where $\llbracket P_i\rrbracket$ is the semantic operator of $P_i$.
We write the alternation $P$ as \begin{equation}\label{g-qgc0}\begin{split}&\mathbf{qif}\ [\overline{q}]:\ |1\rangle\rightarrow P_1\\
&\ \ \square \ \ \ \ \ \ \ \ \ \ |2\rangle\rightarrow P_2\\ &\ \ \ \ \ \ \ \ \ \ \ \ \ \ \ ......\\
&\ \ \square \ \ \ \ \ \ \ \ \ \ |n\rangle\rightarrow P_n\\
&\mathbf{fiq}\end{split}\end{equation}
or
\begin{equation}\label{g-qgc}\mathbf{qif}\ [\overline{q}] (\square i\cdot |i\rangle\rightarrow P_i)\ \mathbf{fiq}\end{equation}
(Whenever the family $\overline{q}$ of quantum variables can be recognised from the context, it can be dropped from the above notation.) The control flow of program in the above alternation is determined by quantum variables $\overline{q}$. For each $1\leq i\leq n$, $P_i$ is guarded by the basis state $|i\rangle$. A superposition of these basis states yields a quantum control flow - superposition of control flows: $$\llbracket P\rrbracket\left(\sum_{i=1}^n\alpha_i |i\rangle|\varphi_i\rangle\right)=\sum_{i=1}^n\alpha_i|i\rangle (\llbracket P_i\rrbracket |\varphi_i\rangle)$$ for all $|\varphi_i\rangle\in\mathcal{H}$ and complex numbers $\alpha_i$ $(1\leq i\leq n)$. This is very different from the classical alternation (\ref{maltn}) or (\ref{maltn1}) of quantum programs where the guards in an alternation cannot be superposed.

Quantum alternation is a convenient notion for describing quantum algorithms; for example, the shift operator of a quantum walk can be written as a quantum alternation: $$S=\mathbf{qif}\ (\square i\cdot |i\rangle\rightarrow S_i)\ \mathbf{fiq}$$ It is interesting to note that even superposition of data can be seen as a special case of superposition of programs: for each $x\in \{0,1\}^n$, let $$V_x:|y\rangle\rightarrow |y\oplus f(x)\rangle$$ for $y=0$ or $1$. Clearly, $V_x$ is a unitary operator on the $2-$dimensional Hilbert space. Then the oracle operator $U_f$ defined in~(\ref{OR}) can be written as a quantum alternation: $$U_f=\mathbf{qif}\ (\square x\in \{0,1\}^n\cdot |x\rangle\rightarrow V_x)\ \mathbf{fiq}$$

\subsubsection{Quantum Choice} Following the idea of defining equation (\ref{wed}) of quantum walk operator $W$, a general form of quantum choice can be easily defined in terms of quantum alternation. Let $P_1,P_2,...,P_n$ be a collection of (quantum) programs, $\overline{q}$ a new family of quantum variables that do not appear in $P_1,P_2,...,P_n$, and $P$ a quantum program acting on $\overline{q}$. Assume that $\{|i\rangle\}$ is an orthonormal basis of the state Hilbert space of the \textquotedblleft coin\textquotedblright\ system denoted by $\overline{q}$. Then the quantum choice of $P_1,P_2,...,P_n$ along $\{|i\rangle\}$ with \textquotedblleft coin-tossing\textquotedblright\ program $P$ is defined as follows:
\begin{equation}\label{quch} [P]
\left(\bigoplus_{i=1}^n\ P_i\right)\stackrel{\triangle}{=}P;\mathbf{qif}\ [\overline{q}] (\square i\cdot |i\rangle\rightarrow P_i)\ \mathbf{fiq}
\end{equation}  Intuitively, quantum choice (\ref{quch}) first runs program $P$ to produce a superposition of the respective execution paths of programs $P_i$ $(1\leq i\leq n)$, and then enters the quantum alternation of $P_1,P_2,...,P_n$ where each $P_i$ is running along its own path within the superposition of paths generated by $P$.

It is interesting to compare quantum choice with probabilistic choice (\ref{proch}). A probabilistic choice is a resolution of nondeterminism where we can simply say that the choice is made according to a certain probability distribution. However, when defining a quantum choice, a \textquotedblleft device\textquotedblright\ that can actually perform the choice, namely a \textquotedblleft quantum coin\textquotedblright, has to be explicitly introduced.

\subsection{Technical Contributions of the Paper}
At the first glance, it seems that the defining equation (\ref{sht}) of shift operator $S$ in Example \ref{qwk} can be smoothly generalized to equation (\ref{aldef}) to define the denotational semantics of a general quantum alternation $P$ of programs $P_1,P_2,...,P_n$. But there is actually a major difficulty in equation (\ref{aldef}). For the case where no quantum measurement occur in any $P_i$ $(1\leq i\leq n)$, the operational semantics of each $P_i$ is simply a sequence of unitary operators, and equation (\ref{aldef}) is not problematic at all. Whenever some $P_i$ contains quantum measurements, however, its semantic structure becomes a tree of linear operators with branching happening at the points where the measurements are performed. Then equation (\ref{aldef}) becomes meaningless within the framework of quantum mechanics, and defining the semantics of quantum alternation $P$ requires to properly combine a collection of trees of quantum operations such that the relevant quantum mechanical principles are still obeyed. This problem will be circumvented in Sections~\ref{sec-comp} and~\ref{Seman} by introducing a semi-classical semantics in terms of operator-valued functions as a middle step toward a purely quantum denotational semantics of programs. Based on this, we systematically develop a theory of quantum programming with quantum alternation and choice. In particular, a set of programming laws for quantum alternation and choice are established.

\subsection{Organisation of the Paper}

We assume that the readers are familiar with the basics of quantum theory including density operator description of mixed quantum states and the super-operator formalism of dynamics of (open) quantum systems; a reader who has no basic knowledge about quantum theory can consult a standard quantum computation textbook~\cite{NC00} or the preliminary sections of several influential quantum programming papers \cite{Se04,SZ00,DP06} and survey \cite{Ga06} as well as the authors' recent papers~\cite{Y11,YF10}.

This paper is organized as follows. A new quantum programming language QGCL is defined in Section \ref{syntax} to support quantum programming with quantum alternation. Section ~\ref{sec-comp} prepares several key ingredients needed in defining the denotational semantics of QGCL, including guarded composition of various quantum operations. The denotational semantics and weakest precondition semantics of QGCL are presented in Section~\ref{Seman}. In Section~\ref{QChoice}, quantum choice is defined in terms of quantum guarded command, and probabilistic choice is implemented by quantum choice through introducing local variables. It should be pointed out that a quantum implementation of probabilistic choice was already given by Zuliani~\cite{Z01} in a different way by using a quantum measurement. A family of algebraic laws for QGCL programs are presented in Section~\ref{Alaws}. Several examples are given in Section \ref{IExam} to illustrate the expressive power of the language QGCL. For readability, some more technical materials are postponed to the appendices. A discussion about the choice of coefficients in the definition of guarded composition of quantum operations is presented in Appendix A. Quantum alternation defined in Sections~\ref{syntax} is guarded by an orthonormal basis of the \textquotedblleft coin\textquotedblright\ space. In Appendix B, we show that the notion of quantum alternation can be generalized to the case where guards are orthogonal subspaces of the \textquotedblleft coin\textquotedblright\ space. All the proofs of lemmas, propositions and theorems are deferred to Appendix C.

\section{QGCL: A Language with Quantum Alternation}\label{syntax}

We first define the syntax of quantum programming language QGCL. It is essentially an extension of Sanders and Zuliani's qGCL \cite{SZ00} obtained by adding quantum alternation. But the presentation of QGCL is quite different from qGCL due to the complications in the semantics of quantum alternation. QGCL also borrows some ideas from Selinger's language QPL~\cite{Se04}.
We assume a countable set $qVar$ of quantum variables ranged over by $q,q_1,q_2,...$. For simplicity of the presentation, we only consider a purely quantum programming language, but we include a countably infinite set $Var$ of classical variables ranged over by $x,y,...$ so that we can use them to record the outcomes of quantum measurements. However, classical computation described by, for example, the assignment statement $x:=e$ in a classical programming language, is excluded.
It is required that the sets of classical and quantum variables are disjoint. For each classical variable $x\in Var$, its type is assumed to be a non-empty set $D_x$; that is, $x$ takes values from $D_x$. In applications, if $x$ is used to store the outcome of quantum measurement $M$, then $spec(M)$ (the set of all possible outcomes of $M$) should be a subset of $D_x$.
For each quantum variable $q\in qVar$, its type is a Hilbert space $type(q)=\mathcal{H}_q$, which is the state space of the quantum system denoted by $q$. For a sequence $\overline{q}=q_1,q_2,\cdots$ of distinct quantum variables, we write: $$type(\overline{q})=\mathcal{H}_{\overline{q}}=\bigotimes_{i\geq 1}\mathcal{H}_{q_i}.$$
So, $type(\overline{q})$ is the state Hilbert space of the composed system denoted by $\overline{q}$.
Similarly, for any set $V\subseteq qVar$, we write: \begin{equation}\label{tpe}type(V)=\mathcal{H}_V=\bigotimes_{q\in V}\mathcal{H}_q\end{equation} for the state Hilbert space of the composed system denoted by $V$. In particular, we write $\mathcal{H}_{all}$  for $type(qVar).$ To simplify the notation, we often identify a sequence of variables with the set of these variables provided they are distinct.

\begin{defn}\label{syn-def} For each QGCL program $P$, we write $var(P)$ for the set of its classical variables£¬ $qvar(P)$ for its quantum variables and $cvar(P)$ for its \textquotedblleft coin\textquotedblright\ variables. Then QGCL programs are inductively defined as follows:\begin{enumerate}
\item $\mathbf{abort}$ and $\mathbf{skip}$ are programs, and $$var(\mathbf{abort})=var(\mathbf{skip})=\emptyset,$$ $$qvar(\mathbf{abort})=qvar(\mathbf{skip})=\emptyset,$$ $$cvar(\mathbf{abort})=cvar(\mathbf{skip})=\emptyset.$$
\item If $\overline{q}$ is a sequence of distinct quantum variables, and $U$ is a unitary operator on $type(\overline{q})$, then $U[\overline{q}]$ is a program, and
$$var(U[\overline{q}])=\emptyset,\ \ \ qvar(U[\overline{q}])=\overline{q}, \ \ \ cvar(U[\overline{q}])=\emptyset.$$
\item If $\overline{q}$ is a sequence of distinct quantum variables, $x$ is a classical variable, $M=\{M_m\}$ is a quantum measurement on $type(\overline{q})$ such that $spec(M)\subseteq D_x,$ where $spec(M)=\{m\}$ is the spectrum of $M$; that is, the set of all possible outcomes of $M$, and $\{P_m\}$ is a family of programs indexed by the outcomes $m$ of measurement $M$ such that $x\notin \bigcup_{m}var(P_m),$ then the classical alternation of $P_m$'s guarded by measurement outcomes $m$'s:
\begin{equation}\label{me-def}P\stackrel{\triangle}{=}\mathbf{measure}\ (\square m\cdot M[\overline{q}:x]=m\rightarrow P_m)\ \mathbf{end}\end{equation} is a program, and $$var(P)=\{x\}\cup\bigcup_mvar(P_m),$$ $$qvar(P)=\overline{q}\cup\bigcup_m qvar(P_m),$$ $$cvar(P)=\bigcup_m cvar(P_m).$$
\item If $\overline{q}$ is a sequence of distinct quantum variables, $\{|i\rangle\}$ is an orthonormal basis of $type(\overline{q})$, and $\{P_i\}$ is a family of programs indexed by the basis states $|i\rangle$'s such that
$$\overline{q}\cap\left(\bigcup_{i} qVar(P_i)\right)=\emptyset,$$ then the quantum alternation of $P_i$'s guarded by basis states $|i\rangle$'s: \begin{equation}\label{ddqa}P\stackrel{\triangle}{=}\mathbf{qif}\ [\overline{q}] \left(\square i\cdot \ |i\rangle\rightarrow P_i\right)\ \mathbf{fiq}\end{equation} is a program, and
$$var(P)=\bigcup_{i} var(P_i),$$ $$qvar(P)=\overline{q}\cup\bigcup_{i} qvar(P_i),$$
$$cvar(P)=\overline{q}\cup\bigcup_{i} cvar(P_i).$$
\item If $P_1$ and $P_2$ are programs such that $var(P_1)\cap var(P_2)=\emptyset$, then $P_1;P_2$ is a program, and $$var(P_1;P_2)=var(P_1)\cup var(P_2),$$ $$qvar(P_1;P_2)=qvar(P_1)\cup qvar(P_2),$$ $$cvar(P_1;P_2)=cvar(P_1)\cup cvar(P_2).$$
\end{enumerate}\end{defn}

The meanings of $\mathbf{abort}$ and $\mathbf{skip}$ are the same as in a classical programming language. Two kinds of statements are introduced in the above definition to describe basic quantum operations, namely unitary transformation and measurement. In the unitary transformation $U[\overline{q}]$, only quantum variables $\overline{q}$ but no classical variables appear, and the transformation is applied to $\overline{q}$. In statement (\ref{me-def}), a measurement $M$ is first performed on quantum variables $\overline{q}$ with the outcome stored in classical variable $x$, and then whenever outcome $m$ is reported, the corresponding subprogram $P_m$ is executed. It is required in statement (\ref{me-def}) that $x\not\in\bigcup_{m}var(P_m)$. This means that the classical variables already used to record the outcomes of the measurements in $P_m$'s are not allowed to store the outcome of a new measurement. This technical requirement is cumbersome, but it can significantly simplify the presentation of the semantics of QGCL. On the other hand, it is not required that the measured quantum variables $\overline{q}$ do not occur in $P_m$. So, measurement $M$ can be performed not only on an external system but also on some quantum variables within $P_m$. The statement (\ref{me-def}) and classical alternation (\ref{maltn}) or (\ref{maltn1}) (of quantum programs) are essentially the same, and the only difference between them is that a classical variable $x$ is added in (\ref{me-def}) to record the measurement outcome.
The intuitive meaning of quantum alternation (\ref{ddqa}) was already carefully explained in Section~\ref{Intro}. Only one thing is worthy to mention: it is required that the variables in $\overline{q}$ do not appear in any $P_i$'s. This indicates that the
\textquotedblleft coin system\textquotedblright\ $\overline{q}$ is external to programs $P_i$'s. Whenever the sequence $\overline{q}$ of quantum variables can be recognized from the context, then it can be dropped from statement (\ref{ddqa}).
The sequential composition $P_1;P_2$ is similar to that in a classical language, and the requirement $var(P_1)\cap var(P_2)=\emptyset$ means that the outcomes of measurements performed at different points are stored in different classical variables. Such a requirement is mainly for technical convenience, and it will considerably simplify the presentation. Obviously, all \textquotedblleft coin\textquotedblright\ are quantum variables: $cvar(P)\subseteq qvar(P)$ for all programs $P$. The set $cvar(P)$ of \textquotedblleft coin\textquotedblright\ variables of program $P$ will be needed in defining a kind of equivalence between quantum programs.
The syntax of QGCL can be summarised as follows:
\begin{equation}\label{qudef}\begin{split}P:=\ &\mathbf{abort}\ |\ \mathbf{skip}\ |\ P_1;P_2\\
 &|\ U[\overline{q}] \ \ \ \ \ \ \ \ \ \ \ \ \ \ \ \ \ \ \ \ \ \ \ \ \ \ \ \ \ \ \ \ \ \ \ \ \ \ \ \ \ \ \ \ \ \ \ \ \ \ \ \ \ \ \ \ \ \ \ \ \ \ \ \ \ \ \ \ \ \ \ \ \ \ \ \ ({\rm unitary\ transformation})\\
&|\ \mathbf{measure}\ (\square m\cdot M[\overline{q}:x]=m\rightarrow P_m)\ \mathbf{end}\ \ \ \ \ \ \ \ \ \ ({\rm classical\ alternation})\\ &|\ \mathbf{qif}\ [\overline{q}] (\square i\cdot |i\rangle\rightarrow P_i)\ \mathbf{fiq}\ \ \ \ \ \ \ \ \ \ \ \ \ \ \ \ \ \ \ \ \ \ \ \ \ \ \ \ \ \ \ \ \ \ \ \ \ \ \ \ \ \ ({\rm quantum\ alternation})
\end{split}\end{equation}

\section{Guarded Compositions of Quantum Operations}\label{sec-comp}

A major difficulty in defining the semantics of QGCL comes from the treatment of quantum alternation. This section provides the key mathematical tool for defining the semantics of quantum alternation, namely guarded composition of quantum operations.

\subsection{Guarded Composition of Unitary Operators}

 To ease the understanding of a general definition of guarded composition, we start with a special case of the guarded composition of unitary operators, which is a straightforward generalisation of the quantum walk shift operator $S$ in Example~\ref{qwk}.

\begin{defn}\label{eqU}For each $1\leq i\leq n$, let $U_i$ be an unitary operator in Hilbert space $\mathcal{H}$. Let $\mathcal{H}_C$ be an auxiliary Hilbert space, called the \textquotedblleft coin space\textquotedblright, with $\{|i\rangle\}$ as an orthonormal basis. Then we define a linear operator: $$U\stackrel{\triangle}{=}\square_{i=1}^n\ |i\rangle\rightarrow U_i$$ in $\mathcal{H}_C\otimes\mathcal{H}$ by \begin{equation}\label{defcu}U(|i\rangle|\psi\rangle)=|i\rangle(U_i|\psi\rangle)\end{equation} for any $|\psi\rangle\in\mathcal{H}$ and for any $1\leq i\leq n$. Then by linearity we have: \begin{equation}\label{eqUU}U\left(\sum_{i,j}\alpha_{ij}|i\rangle|\psi_j\rangle\right)=\sum_{i,j}\alpha_{ij}|i\rangle\left(U_i|\psi_j\rangle\right)\end{equation} for any $|\psi_j\rangle\in\mathcal{H}$ and complex numbers $\alpha_{ij}$. The operator $U$ is called the guarded composition of $U_i$ $(1\leq i\leq n)$ along the basis $\{|i\rangle\}$.\end{defn}

A routine calculation yields:

\begin{lem}\begin{enumerate}\item The guarded composition $\square_{i}\ |i\rangle\rightarrow U_i$ is an unitary operator in $\mathcal{H}_C\otimes\mathcal{H}$. \item For any two orthonormal basis $\{|i\rangle\}$ and $\{|\varphi_i\rangle\}$ of the \textquotedblleft coin space\textquotedblright\ $\mathcal{H}_C$, there exists an unitary operator $U_C$ such that $|\varphi_i\rangle=U_C|i\rangle$ for all $i$, and the two compositions along different bases $\{|i\rangle\}$ and $\{|\varphi_i\rangle\}$ are related to each other by $$\square_i|\varphi_i\rangle\rightarrow U_i=(U_C\otimes I_\mathcal{H})(\square_i|i\rangle\rightarrow U_i)(U_C^\dag\otimes I_\mathcal{H})$$ where $I_\mathcal{H}$ is the identity operator in $\mathcal{H}$.\end{enumerate}
\end{lem}

Clause (1) of the above lemma indicates that the guarded composition of unitary operators is well defined, and clause (2) shows that the choice of orthonormal basis of the \textquotedblleft coin space\textquotedblright\ is not essential for the definition of guarded composition.

The guarded composition of unitary operators is nothing new; it is just a quantum multiplexor introduced in \cite{SBM06} as a useful tool in the synthesis of quantum logic circuits.

\begin{exam} A quantum multiplexor (QMUX for short) is a quantum generalisation of multiplexor, a well-known notion in digit logic. A QMUX $U$ with $k$ select qubits and $d-$qubit-wide data bus can be represented by a block-diagonal matrix:
$$U=diag (U_0,U_1,...,U_{2^k-1})=\left(\begin{array}{cccc}U_0& & & \\
& U_1& & \\ & & ... & \\ & & & U_{2^k-1}
\end{array}\right).$$ Multiplexing $U_0,U_1,...,U_{2^k-1}$ with $k$ select qubits is exactly the guarded composition $$\square_{i=0}^{2^k-1}|i\rangle\rightarrow U_i$$ along the computational basis $\{|i\rangle\}$ of $k$ qubits.\end{exam}

\subsection{Operator-Valued Functions}

 A general form of guarded composition of quantum operations cannot be defined by a straightforward generalization of Definition~\ref{eqU}. Instead, we need an auxiliary notion of operator-valued function. For any Hilbert space $\mathcal{H}$, we write $\mathcal{L}(\mathcal{H})$ for the space of (bounded linear) operators in $\mathcal{H}$.

\begin{defn} Let $\Delta$ be a nonempty set. Then a function $F:\Delta\rightarrow \mathcal{L}(\mathcal{H})$ is called an operator-valued function in $\mathcal{H}$ over $\Sigma$ if \begin{equation}\label{sem-con}\sum_{\delta\in \Delta}F(\delta)^\dag\cdot F(\delta)\sqsubseteq I_{\mathcal{H}},\end{equation} where $I_\mathcal{H}$ is the identity operator in $\mathcal{H}$, and $\sqsubseteq$ stands for the L\"{o}wner order; that is, $A\sqsubseteq B$ if and only if $B-A$ is a positive operator. In particular, $F$ is said to be full when equation~(\ref{sem-con}) becomes equality.
\end{defn}

The simplest examples of operator-valued function are unitary operators and quantum measurements.

\begin{exam}\begin{enumerate}\item A unitary operator $U$ in Hilbert space $\mathcal{H}$ can be seen as a full operator-valued function over a singleton $\Delta=\{\epsilon\}$. This function maps $\epsilon$ to $U$.
\item A quantum measurement $M=\{M_m\}$ in Hilbert space $\mathcal{H}$ can be seen as a full operator-valued function over its spectrum $spec(M)=\{m\}$ (the set of possible measurement outcomes). This function maps each measurement outcome $m$ to the corresponding measurement operator $M_m$.
\end{enumerate}\end{exam}

More generally, a super-operator (or quantum operation) defines a family of operator-valued functions. Let $\mathcal{E}$ be a super-operator in Hilbert space $\mathcal{H}$. Then $\mathcal{E}$ has the Kraus operator-sum representation: $$\mathcal{E}=\sum_i E_i\circ E_i^\dag,$$ meaning: $$\mathcal{E}(\rho)=\sum_i E_i\rho E_i^\dag$$ for all density operators $\rho$ in $\mathcal{H}$ (see \cite{NC00}, Chapter 8). For such a representation, we set $\Delta=\{i\}$ for the set of indexes, and define an operator-valued function over $\Delta$ by $$F(i)=E_i$$ for every $i$. Since operator-sum representation of $\mathcal{E}$ is not unique, $\mathcal{E}$ defines not only a single operator-valued function. We write $\mathbb{F}(\mathcal{E})$ for the family of operator-valued functions defined by all Kraus operator-sum representations of $\mathcal{E}$. Conversely, an operator-valued function determines uniquely a super-operator.

\begin{defn}\label{sfop} Let $F$ be an operator-valued function in Hilbert space $\mathcal{H}$ over set $\Delta$. Then $F$ defines a super-operator $\mathcal{E}(F)$ in $\mathcal{H}$ as follows: $$\mathcal{E}(F)=\sum_{\delta\in\Delta} F(\delta)\circ F(\delta)^\dag;$$ that is, $$\mathcal{E}(F)(\rho)=\sum_{\delta\in\Delta} F(\delta)\rho F(\delta)^\dag$$ for every density operator $\rho$.
\end{defn}

For a family $\mathbb{F}$ of operator-valued functions, we write: $$\mathcal{E}(\mathbb{F})=\{\mathcal{E}(F):F\in\mathbb{F}\}.$$ It is obvious that $\mathcal{E}(\mathbb{F}(\mathcal{E}))=\{\mathcal{E}\}$ for each super-operator $\mathcal{E}$. On the other hand, for any operator-valued function $F$ over $\Delta=\{\delta_1,...,\delta_k\}$, it follows from Theorem 8.2 in \cite{NC00} that $\mathbb{F}(\mathcal{E}(F))$ consists of all operator-valued functions $G$ over some $\Gamma=\{\gamma_1,...,\gamma_l\}$ such that $$G(\gamma_i)=\sum_{j=1}^n u_{ij}\cdot F(\delta_j)$$ for each $1\leq i\leq n$, where $n=\max (k,l)$, $U=(u_{ij})$ is an $n\times n$ unitary matrix, $F(\delta_i)=G(\gamma_j)=0_\mathcal{H}$ for all $k+1< i\leq n$ and $l+1< j\leq n$, and $0_\mathcal{H}$ is the zero operator in $\mathcal{H}$.

\subsection{Guarded Composition of Operator-Valued Functions}

We need to introduce a notation before defining guarded composition of operator-valued functions. Let $\Delta_i$ be a nonempty set for every $1\leq i\leq n$. Then the superposition of $\Delta_i$ $(1\leq i\leq n)$ is defined as follows:
\begin{equation}\label{Ssup}\bigoplus_{i=1}^n\Delta_i=\{\oplus_{i=1}^n\delta_i:\delta_i\in\Delta_i\ {\rm for\ every}\ 1\leq i\leq n\}.\end{equation} Here, $\oplus_{i=1}^n\delta_i$ is simply a notation indicating a combination of $\delta_i$ $(1\leq i\leq n)$, and we do not need to care its meaning.

\begin{defn}\label{qgu}For each $1\leq i\leq n$, let $F_i$ be an operator-valued function in Hilbert space $\mathcal{H}$ over set $\Delta_i$. Let $\mathcal{H}_C$ be a \textquotedblleft coin\textquotedblright\ Hilbert space with $\{|i\rangle\}$ as an orthonormal basis. Then the guarded composition
$$F\stackrel{\triangle}{=}\square_{i=1}^n\ |i\rangle\rightarrow F_i$$
of $F_i$ $(1\leq i\leq n)$ along the basis $\{|i\rangle\}$ is defined to be the operator-valued function
$$F:\bigoplus_{i=1}^n\Delta_i\rightarrow \mathcal{L}(\mathcal{H}_C\otimes\mathcal{H})$$ in $\mathcal{H}_C\otimes\mathcal{H}$ over $\bigoplus_{i=1}^n\Delta_i$. For any $\delta_i\in\Delta_i$ $(1\leq i\leq n)$, $F(\oplus_{i=1}^n\delta_i)$ is an operator in $\mathcal{H}_C\otimes\mathcal{H}$ defined as follows: for each $|\Psi\rangle\in\mathcal{H}_C\otimes\mathcal{H}$, there is a unique tuple $(|\psi_1\rangle, ..., |\psi_n\rangle)$ such that $|\psi_1\rangle, ..., |\psi_n\rangle\in\mathcal{H}$ and $|\Psi\rangle$ can be written as $$|\Psi\rangle=\sum_{i=1}^n|i\rangle|\psi_i\rangle,$$ and then we define
\begin{equation}\label{coef0}F(\oplus_{i=1}^n\delta_i)|\Psi\rangle=\sum_{i=1}^n\left(\prod_{k\neq i}\lambda_{k\delta_k}\right)|i\rangle(F_i(\delta_i)|\psi_i\rangle)\end{equation} where \begin{equation}\label{coef1}\lambda_{k\delta_k}=\sqrt{\frac{tr F_k(\delta_k)^\dag F_{k}(\delta_k)}{\sum_{\tau_k\in\Delta_k}tr F_{k}(\tau_k)^\dag F_{k}(\tau_k)}}.\end{equation} In particular, if $F_k$ is full and $d=\dim\mathcal{H}<\infty$, then $$\lambda_{k\delta_k}=\sqrt{\frac{tr F_{k}(\delta_k)^\dag F_{k}(\delta_k)}{d}}$$ for any $\delta_k\in\Delta_k$ $(1\leq k\leq n)$.
\end{defn}

Intuitively, the square $\lambda_{k\delta_k}^2$ of the coefficients defined in equation~(\ref{coef1}) can be understood as a kind of conditional probability. A further discussion on the choice of coefficients in equation~(\ref{coef1}) is given in Appendix A. The following lemma shows that the guarded composition of operator-valued functions is well-defined, and the choice of orthonormal basis of the \textquotedblleft coin space\textquotedblright\ is not essential in its definition.

\begin{lem}\label{gulem}\begin{enumerate}\item The guarded composition $F\stackrel{\triangle}{=}\square_{i=1}^n\ |i\rangle\rightarrow F_i$ is an operator-valued function in $\mathcal{H}_C\otimes\mathcal{H}$ over $\bigoplus_{i=1}^n\Delta_i$. In particular, if all $F_i$ $(1\leq i\leq n)$ are full, then so is $F$.
\item For any two orthonormal bases $\{|i\rangle\}$ and $\{|\varphi_i\rangle\}$ of the \textquotedblleft coin space\textquotedblright\ $\mathcal{H}_C$, there exists an unitary operator $U_C$ such that $|\varphi_i\rangle= U_C|i\rangle$ for all $i$, and the two compositions along different bases $\{|i\rangle\}$ and $\{|\varphi_i\rangle\}$ are related to each other by $$\square_{i=1}^n |\varphi_i\rangle\rightarrow F_i=(U_C\otimes I_\mathcal{H})\cdot (\square_{i=1}^n|i\rangle\rightarrow F_i)\cdot (U_C^\dag\otimes I_\mathcal{H});$$ that is, $$(\square_{i=1}^n |\varphi_i\rangle\rightarrow F_i)(\oplus_{i=1}^n\delta_i)=(U_C\otimes I_\mathcal{H})(\square_{i=1}^n|i\rangle\rightarrow F_i)(\oplus_{i=1}^n\delta_i)(U_C^\dag\otimes I_\mathcal{H})$$ for any $\delta_1\in\Delta_1,...,\delta_n\in\Delta_n.$\end{enumerate}
\end{lem}

It is easy to see that whenever $\Delta_i$ is a singleton for all $1\leq i\leq n$, then all $\lambda_{k\delta_k}=1$ and equation~(\ref{coef0}) degenerates to~(\ref{eqUU}). So, the above definition is a generalisation of guarded composition of unitary operators introduced in Definition~\ref{eqU}. On the other hand, it can also be used to compose quantum measurements as shown in the following simple example.

\begin{exam}\label{gmeas} We consider a guarded composition of two simplest quantum measurements. Let $M^{(0)}$ be the measurement on a qubit (the principal qubit) $q$ in the computational basis $|0\rangle, |1\rangle$ , i.e. $M^{(0)}=\{M_0^{(0)},M_1^{(0)}\}$, where $M_0^{(0)}=|0\rangle\langle 0|,$ $M_1^{(0)}=|1\rangle\langle 1|$, and let $M^{(1)}$ be the measurement of the same qubit but in a different basis: $$|\pm\rangle=\frac{1}{\sqrt{2}}(|0\rangle\pm |1\rangle),$$ i.e. $M^{(1)}=\{M_+^{(1)},M_-^{(1)}\}$, where $M_+^{(1)}=|+\rangle\langle +|,M_-^{(1)}=|-\rangle\langle -|$. Then the guarded composition of $M^{(0)}$ and $M^{(1)}$ along the computational basis of another qubit (the \textquotedblleft coin qubit\textquotedblright)\ $q_C$ is the measurement \begin{equation*}M=(|0\rangle\rightarrow M^{(0)})\ \square\ (|1\rangle\rightarrow M^{(1)})=\{M_{0+}, M_{0-}, M_{1+}, M_{1-}\}\end{equation*} on two qubits $q$ and $q_C$, where $ij$ is an abbreviation of $i\oplus j$, and
$$M_{ij}(|0\rangle_{q_C}|\psi_0\rangle_q+|1\rangle_{q_C}|\psi_1\rangle_q)=\frac{1}{\sqrt{2}}(|0\rangle_{q_C} M_i^{(0)}|\psi_0\rangle_q+|1\rangle_{q_C} M_j^{(1)}|\psi_1\rangle_q)$$ for any states $|\psi_0\rangle, |\psi_1\rangle$ of the principal qubit $q$ and $i\in\{0,1\}, j\in\{+,-\}$. Furthermore, for each state $|\Psi\rangle$ of two qubits $q, q_C$ and for any $i\in\{0,1\}, j\in\{+,-\}$, a routine calculation yields that the probability that the outcome is $ij$ when performing the guarded composition $M$ of $M^{(0)}$ and $M^{(1)}$ on the two qubit system $q_Cq$ in state $|\Psi\rangle$ is  $$p(i,j||\Psi\rangle,M)=\frac{1}{2}\left[p(i|_{q_C}\langle 0|\Psi\rangle,M^{(0)})+p(j|_{q_C}\langle 1|\Psi\rangle,M^{(1)})\right],$$ where: \begin{enumerate}\item if $|\Psi\rangle=|0\rangle_{q_C}|\psi_0\rangle_q+|1\rangle_{q_C}|\psi_1\rangle_q$, then $_{q_C}\langle k|\Psi\rangle=|\psi_k\rangle_q$ is the \textquotedblleft conditional\textquotedblright\ state of the principal qubit $q$ given that the two qubit system $q_Cq$ is in state $|\Psi\rangle$ and the \textquotedblleft coin\textquotedblright\ qubit $q_C$ is in the basis state $|k\rangle$ for $k=0,1$;
\item $p(i|_{q_C}\langle 0|\Psi\rangle,M^{(0)})$ is the probability that the outcome is $i$ when performing measurement $M^{(0)}$ on qubit $q$ in state $_{q_C}\langle 0|\Psi\rangle$; \item $p(j|_{q_C}\langle 1|\Psi\rangle,M^{(1)})$ is the probability that the outcome is $j$ when performing measurement $M^{(1)}$ on qubit $q$ in state $_{q_C}\langle 1|\Psi\rangle$.
\end{enumerate}
\end{exam}

\subsection{Guarded Composition of Super-Operators}

Now the guarded composition of a family of super-operators can be defined through the guarded composition of the operator-valued functions generated from them.

\begin{defn}\label{def-gsup}For each $1\leq i\leq n$, let $\mathcal{E}_i$ be a super-operator in Hilbert space $\mathcal{H}$. Let $\mathcal{H}_C$ be a \textquotedblleft coin\textquotedblright\ Hilbert space with $\{|i\rangle\}$ as an orthonormal basis. Then the guarded composition of $\mathcal{E}_i$ $(1\leq i\leq n)$ along the basis $\{|i\rangle\}$ is defined to be the family of super-operators in $\mathcal{H}_C\otimes\mathcal{H}$:
\begin{equation*}\square_{i=1}^n\ |i\rangle\rightarrow\mathcal{E}_i=\{\mathcal{E}(\square_{i=1}^n\ |i\rangle\rightarrow F_i): F_i\in\mathbb{F}(\mathcal{E}_i)\ {\rm for\ every}\ 1\leq i\leq n\},\end{equation*} where $\mathbb{F}(\mathcal{F})$ stands for the family of operator-valued functions defined by all Kraus operator-sum representations of an super-operator $\mathcal{F}$, and $\mathcal{E}(F)$ is the super-operator defined by an operator-valued function $F$ (see Definition \ref{sfop}).
\end{defn}

It is easy to see that if $n=1$ then the above guarded composition of super-operators consists of only $\mathcal{E}_1$. For $n>1$, however,  it is usually not a singleton, as shown by the following:

\begin{exam}\label{eexgd} For any unitary operator $U$ in a Hilbert space $\mathcal{H}$, we write $\mathcal{E}_U=U\circ U^\dag$ for the super-operator defined by $U$; that is, $$\mathcal{E}_U(\rho)=U\rho U^\dag$$ for all density operators $\rho$ in $\mathcal{H}$. Now suppose that $U_0$ and $U_1$ are two unitary operators in $\mathcal{H}$. Let $U$ be the composition of $U_0$ and $U_1$ guarded by the computational basis $|0\rangle, |1\rangle$ of a qubit: $$U=(|0\rangle\rightarrow U_0)\square (|1\rangle\rightarrow U_1).$$ Then $\mathcal{E}_U$ is an element of the guarded composition $$\mathcal{E}=(|0\rangle\rightarrow \mathcal{E}_{U_0})\square (|1\rangle\rightarrow \mathcal{E}_{U_1})$$ of super-operators $\mathcal{E}_{U_0}$ and $\mathcal{E}_{U_1}.$ But $\mathcal{E}$ contains more than one element. Indeed, it holds that
$$\mathcal{E}=\{\mathcal{E}_{U_\theta}=U_\theta\circ\ U_\theta^\dag:0\leq \theta<2\pi\},$$ where
$$U_\theta=(|0\rangle\rightarrow U_0)\square (|1\rangle\rightarrow e^{i\theta} U_1).$$
The non-uniqueness of the members of the guarded composition $\mathcal{E}$ is caused by the relative phase $\theta$ between $U_0$ and $U_1$.
\end{exam}

 For any two super-operators $\mathcal{E}_1$ and $\mathcal{E}_2$ in a Hilbert space $\mathcal{H}$, their sequential composition $\mathcal{E}_2;\mathcal{E}_1$ is the super-operator in $\mathcal{H}$ defined by $$(\mathcal{E}_2;\mathcal{E}_1)(\rho)=\mathcal{E}_2(\mathcal{E}_1(\rho))$$ for any density operator $\rho$ in $\mathcal{H}$. For any super-operator $\mathcal{E}$ and any set $\Omega$ of super-operators in Hilbert space $\mathcal{H}$, we define the sequential composition of $\Omega$ and $\mathcal{E}$ by $$\mathcal{E};\Omega=\{\mathcal{E};\mathcal{F}:\mathcal{F}\in\Omega\}.$$ The following lemma can be easily derived from Lemma~\ref{gulem} (2), and it shows that the choice of orthonormal basis of the \textquotedblleft coin space\textquotedblright\ is not essential for the guarded composition of super-operators.

\begin{lem}For any two orthonormal bases $\{|i\rangle\}$ and $\{|\varphi_i\rangle\}$ of the \textquotedblleft coin space\textquotedblright\ $\mathcal{H}_C$, there exists an unitary operator $U_C$ such that $|\varphi_i\rangle=U_C|i\rangle$ for all $i$, and the two compositions along different bases $\{|i\rangle\}$ and $\{|\varphi_i\rangle\}$ are related to each other by $$\square_{i=1}^n |\varphi_i\rangle\rightarrow \mathcal{E}_i=\mathcal{E}_{U_C\otimes I_\mathcal{H}}; \left[(\square_{i=1}^n|i\rangle\rightarrow \mathcal{E}_i);\mathcal{E}_{U_C^\dag\otimes I_{\mathcal{H}}}\right],$$ where $\mathcal{E}_{U_C\otimes I_\mathcal{H}}$ and $\mathcal{E}_{U_C^\dag \otimes I_\mathcal{H}}$ are the super-operators in $\mathcal{H}_C\otimes\mathcal{H}$ defined by unitary operators $U_C\otimes I_\mathcal{H}$ and $U_C^\dag\otimes I_\mathcal{H}$, respectively.\end{lem}

\section{Semantics of QGCL}\label{Seman}

With the preparation in Section \ref{sec-comp}, we are ready to define the semantics of language QGCL. We first introduce several notations needed in this section. Let $\mathcal{H}$ and $\mathcal{H}^\prime$ be two Hilbert spaces, and let $E$ be an operator in $\mathcal{H}$. Then the cylindrical extension of $E$ in $\mathcal{H}\otimes\mathcal{H}^\prime$ is defined to be the operator $E\otimes I_{\mathcal{H}^\prime}$, where $I_{\mathcal{H}^\prime}$ is the identity operator in $\mathcal{H}^\prime$. For simplicity, we will write $E$ for $E\otimes I_{\mathcal{H}^\prime}$ whenever confusion does not happen. Let $F$ be an operator-valued function in $\mathcal{H}$ over $\Delta$. Then the cylindrical extension of $F$ in $\mathcal{H}\otimes\mathcal{H}^\prime$ is the operator-valued function $\overline{F}$ in $\mathcal{H}\otimes\mathcal{H}^\prime$ over $\Delta$ defined by $$\overline{F}(\delta)=F(\delta)\otimes I_{\mathcal{H}^\prime}$$ for every $\delta\in\Delta$. For simplicity, we often write $F$ for $\overline{F}$ whenever confusion can be excluded from the context.
Furthermore, let $\mathcal{E}=\sum_i E_i\circ E_i^\dag$ be a super-operator in $\mathcal{H}$. Then the cylindrical extension $\overline{\mathcal{E}}$ of $\mathcal{E}$ in $\mathcal{H}\otimes\mathcal{H}^\prime$ is defined to be the super-operator: $$\overline{\mathcal{E}}=\sum_i(E_i\otimes I_{\mathcal{H}^\prime})\circ (E_i^\dag\otimes I_{\mathcal{H}^\prime}).$$ For simplicity, $\mathcal{E}$ will be used to denote its extension $\overline{\mathcal{E}}$ when no confusion occurs. In particular, if $E$ is an operator in $\mathcal{H}$, and $\rho$ is a density operator in $\mathcal{H}\otimes\mathcal{H}^\prime$, then $E\rho E^\dag$ should be understood as $(E\otimes I_{\mathcal{H}^\prime})\rho(E^\dag\otimes I_{\mathcal{H}^\prime})$.

\subsection{Classical States}

We now define the states of classical variables in QGCL. As already stated in Section~\ref{syntax}, they will be only used to record the outcomes of quantum measurements.

\begin{defn}\label{cl-state}The (partial) classical states and their domains are inductively defined as follows:\begin{enumerate} \item $\epsilon$ is a classical state, called the empty state, and $dom(\epsilon)=\emptyset$; \item If $x\in Var$ is a classical variable, and $a\in D_x$ is an element of the domain of $x$, then $[x\leftarrow a]$ is a classical state, and $dom([x\leftarrow a])=\{x\}$; \item If both $\delta_1$ and $\delta_2$ are classical states, and $dom(\delta_1)\cap dom(\delta_2)=\emptyset$, then $\delta_1\delta_2$ is a classical state, and $dom(\delta_1\delta_2)=dom(\delta_1)\cup dom(\delta_2)$; \item If $\delta_i$ is a classical state for every $1\leq i\leq n$, then $\oplus_{i=1}^n\delta_i$ is a classical state, and $$dom(\oplus_{i=1}^n\delta_i)=\bigcup_{i=1}^n dom(\delta_i).$$
\end{enumerate}
\end{defn}

Intuitively, a classical state $\delta$ defined by clauses (1) to (3) in the above definition can be seen as a (partial) assignment to classical variables; more precisely, $\delta$ is an element of $\prod_{x\in dom(\delta)}D_x;$ that is, a choice function: $$\delta: dom(\delta)\rightarrow\bigcup_{x\in dom(\delta)}D_x$$ such that $\delta(x)\in D_x$ for every $x\in dom(\delta)$. In particular, $\epsilon$ is the empty function. Since $\prod_{x\in\emptyset}D_x=\{\epsilon\},$ $\epsilon$ is the only possible state with empty domain. The state $[x\leftarrow a]$ assigns value $a$ to variable $x$ but the values of the other variables are undefined. The composed state $\delta_1\delta_2$ can be seen as the assignment to variables in $dom(\delta_1)\cup dom(\delta_2)$ given by
\begin{equation}\label{comp-sta}(\delta_1\delta_2)(x)=\begin{cases}\delta_1(x) &{\rm if}\ x\in dom(\delta_1),\\ \delta_2(x) &{\rm if}\ x\in dom(\delta_2).\end{cases}\end{equation}
Equation~(\ref{comp-sta}) is well-defined since it is required that $dom(\delta_1)\cap dom(\delta_2)=\emptyset.$ In particular, $\epsilon\delta=\delta\epsilon=\delta$ for any state $\delta$, and
if $x\notin dom(\delta)$ then $\delta [x\leftarrow a]$ is the assignment to variables in $dom(\delta)\cup\{x\}$ given by
$$\delta[x\leftarrow a](y)=\begin{cases}
\delta(y) &{\rm if}\ y\in dom(\delta),\\ a &{\rm if}\ y=x.
\end{cases}$$ Hence, $[x_1\leftarrow a_1]\cdots [x_n\leftarrow a_n]$ is a classical state that assigns value $a_i$ to variable $x_i$ for all $1\leq i\leq n$. It will be abbreviated to $[x_1\leftarrow a_1,\cdots,x_n\leftarrow a_n]$ in the sequel. The state $\oplus_{i=1}^n\delta_i$ defined by clause (4) in Definition~\ref{cl-state} can be thought of as a kind of superposition of $\delta_i$ $(1\leq i\leq n)$. It will be used in defining the semantics of quantum alternation.

\subsection{Semi-Classical Denotational Semantics}

The semi-classical semantics of QGCL is a step stone for defining its purely quantum semantics. For each QGCL program $P$, we write $\Delta(P)$ for the set of all possible states of its classical variables. The semi-classical denotational semantics $\lceil P\rceil$ of $P$ will be defined as an operator-valued function in $\mathcal{H}_{qvar(P)}$ over $\Delta(P)$, where $\mathcal{H}_{qvar(P)}$ is the type of quantum variables occurring in $P$. In particular, if $qvar(P)=\emptyset$; for example $P=\mathbf{abort}$ or $\mathbf{skip}$, then $\mathcal{H}_{qvar(P)}$ is a one-dimensional space $\mathcal{H}_\emptyset$, and an operator in $\mathcal{H}_\emptyset$ can be identified with a complex number; for instance the zero operator is number $0$ and the identity operator is number $1$. For any set $V\subseteq qVar$ of quantum variables, we write $I_V$ for the identity operator in Hilbert space $\mathcal{H}_V$ (see equation (\ref{tpe})).

\begin{defn}\label{defsem}The classical states $\Delta(P)$ and semi-classical semantic function $\lceil P\rceil$ of a QGCL program $P$ are inductively defined as follows:\begin{enumerate} \item $\Delta(\mathbf{abort})=\{\epsilon\}$, and $\lceil\mathbf{abort}\rceil(\epsilon)=0$; \item $\Delta(\mathbf{skip})=\{\epsilon\}$, and $\lceil\mathbf{skip}\rceil(\epsilon)=1;$ \item $\Delta(U[\overline{q}])=\{\epsilon\}$, and $\lceil U[\overline{q}]\rceil(\epsilon) =U_{\overline{q}},$ where $U_{\overline{q}}$ is the unitary operator $U$ acting in $\mathcal{H}_{\overline{q}}$;
\item If $P$ is a classical alternation: $$P\stackrel{\triangle}{=}\mathbf{measure}\ \left(\square m\cdot M[\overline{q}:x]=m\rightarrow P_m\right)\ \mathbf{end},$$ where quantum measurement $M=\{M_m\}$, then $$\Delta(P)=\bigcup_m\{\delta[x\leftarrow m]:\delta\in D(P_m)\},$$ $$\lceil P\rceil (\delta[x\leftarrow m])=(\lceil P_m\rceil (\delta)\otimes I_{V\setminus qvar(P_m)})\cdot (M_m\otimes I_{V\setminus\overline{q}})$$ for every $\delta\in \Delta(P_m)$ and for every outcome $m$, where $V=\overline{q}\cup\bigcup_{m}qvar(P_m)$; \item If $P$ is a quantum alternation: $$P\stackrel{\triangle}{=}\mathbf{qif}\ [\overline{q}] \left(\square i\cdot |i\rangle\rightarrow P_i\right)\ \mathbf{fiq},$$ then $$\Delta(P)=\bigoplus_{i}\Delta(P_i),$$ \begin{equation}\label{ddalt}\lceil P\rceil =\square_{i}\ |i\rangle\rightarrow \lceil P_i\rceil,\end{equation} where operation $\bigoplus$ is defined by equation~(\ref{Ssup}), and $\square$ in equation (\ref{ddalt}) stands for the guarded composition of operator-valued functions (see Definition~\ref{qgu});
\item If $P=P_1;P_2$, then \begin{equation}\label{sem-seq}\Delta (P)=\Delta(P_1);\Delta(P_2)=\{\delta_1\delta_2:\delta_1\in\Delta(P_1)\ {\rm and}\ \delta_2\in\Delta(P_2)\},\end{equation}
\begin{equation*}\lceil P\rceil (\delta_1\delta_2)=(\lceil P_2\rceil (\delta_2)\otimes I_{V\setminus qvar(P_2)})\cdot (\lceil P_1\rceil (\delta_1)\otimes I_{V\setminus qvar(P_1)})\end{equation*} where $V=qvar(P_1)\cup qvar(P_2)$;
\end{enumerate}
\end{defn}

 Intuitively, if a quantum program $P$ does not contain any quantum alternation, then its semantic structure can be seen as a tree with its nodes labelled by basic commands and its edges by linear operators. This tree grows up from the root in the following way: if the current node is labelled by a unitary transformation $U$, then a single edge stems from the node and it is labelled by $U$; and if the current node is labelled by a measurement $M=\{M_m\}$, then for each possible outcome $m$, an edge stems from the node and it is labelled by the corresponding measurement operator $M_m$. Obviously, branching in the semantic tree comes from the different possible outcomes of a measurement in $P$. Each classical state $\delta\in\Delta(P)$ is corresponding to a branch in the semantic tree of $P$, and it denotes a possible path of execution. Furthermore, the value of semantic function $\lceil P\rceil$ in state $\delta$ is the (sequential) composition of the operators labelling the edges of $\delta$. This can be clearly seen from clauses (3), (4) and (6) of the above definition. Since it is required in Definition~\ref{syn-def} that $var(P_1)\cap var(P_2)=\emptyset$ in the sequential composition $P_1;P_2$, we have $dom(\delta_1)\cap dom(\delta_2)=\emptyset$ for any $\delta_1\in\Delta(P_1)$ and $\delta_2\in\Delta(P_2)$. Thus, equation~(\ref{sem-seq}) is well-defined.

 The semantic structure of a quantum program $P$ with quantum alternations is much more complicated. We can imagine it as a tree with superpositions of nodes that generate superpositions of branches. The value of semantic function $\lceil P\rceil$ in a superposition of branches is then defined as the guarded composition of the values in these branches.

\subsection{Purely Quantum Denotational Semantics}

Now the purely quantum semantics of a quantum program can be naturally defined as the super-operator induced by its semi-classical semantic function.

\begin{defn}\label{den-def}For each QGCL program $P$, its purely quantum denotational semantics is the super-operator $\llbracket P\rrbracket$ in $\mathcal{H}_{qvar(P)}$ defined as follows:
\begin{equation}\label{ddpq}\llbracket P\rrbracket =\mathcal{E}(\lceil P\rceil)=\sum_{\delta\in \Delta(P)}\lceil P\rceil (\delta)\circ \lceil P\rceil(\delta)^\dag.\end{equation}
\end{defn}

The following proposition presents a representation of the purely quantum semantics of a program in terms of its subprograms.

\begin{prop}\label{qsem}\begin{enumerate}
\item $\llbracket\mathbf{abort}\rrbracket =0;$
\item $\llbracket\mathbf{skip} \rrbracket =1;$
\item $\llbracket P_1;P_2\rrbracket  =\llbracket P_1\rrbracket ;\llbracket P_2\rrbracket;$
\item $\llbracket U[\overline{q}]\rrbracket = U_{\overline{q}}\circ U_{\overline{q}}^\dag;$
\item $$\llbracket\mathbf{measure}\ \left(\square m\cdot M[\overline{q}:x]=m\rightarrow P_m\right)\ \mathbf{end}\rrbracket =\sum_m \left[(M_m\circ M_m^\dag);\llbracket P_m\rrbracket\right].$$
    Here, $\llbracket P_m\rrbracket$ should be seen as a cylindrical extension in $\mathcal{H}_{V}$ from $\mathcal{H}_{qvar(P_m)}$, $M_m\circ M_m^\dag$ is seen as a cylindrical extension in $\mathcal{H}_{V}$ from $\mathcal{H}_{\overline{q}}$, and $V=\overline{q}\cup\bigcup_{m} qvar(P_m)$;
\item \begin{equation}\label{ppalt}\llbracket\mathbf{qif}\ [\overline{q}] \left(\square i\cdot |i\rangle\rightarrow P_i\right)\ \mathbf{fiq}\rrbracket\in \square_{i}\ |i\rangle\rightarrow \llbracket P_i\rrbracket.\end{equation}
    Here $\llbracket P_i\rrbracket$ should be understood as a cylindrical extension in $\mathcal{H}_{V}$ from $\mathcal{H}_{qvar(P_i)}$ for every $1\leq i\leq n$, and $V=\overline{q}\cup\bigcup_{i} qvar(P_i)$.\end{enumerate}\end{prop}

The above proposition shows that the purely quantum denotational semantics is almost compositional, but it is not completely compositional because the symbol \textquotedblleft$\in$\textquotedblright\ appears in equation (\ref{ppalt}) of the above proposition. The symbol \textquotedblleft$\in$\textquotedblright\ can be understood as a refinement relation. It is worth noting that in general \textquotedblleft$\in$" cannot be replaced by equality. This is exactly the reason that the purely quantum semantics of a program has to be derived through its semi-classical semantics but cannot be defined directly by a structural induction.

It should be stressed that the symbol \textquotedblleft$\in$\textquotedblright\ in equation (\ref{ppalt}) does not mean that the purely quantum denotational semantics of quantum alternation \textquotedblleft$\mathbf{qif}\ [\overline{q}] \left(\square i\cdot |i\rangle\rightarrow P_i\right)\ \mathbf{fiq}$\textquotedblright\ is not well-defined. In fact, it is uniquely defined by equations (\ref{ddalt}) and (\ref{ddpq}) as a super-operator. The right-hand side of equation (\ref{ppalt}) is not the denotational semantics of any program. It is the guarded composition of the denotational semantics of programs $P_i$. Since it is the guarded composition of a family of super-operators, it can be a set consisting of more than one super-operator, as shown in Example \ref{eexgd}. The semantics of quantum alternation \textquotedblleft$\mathbf{qif}\ [\overline{q}] \left(\square i\cdot |i\rangle\rightarrow P_i\right)\ \mathbf{fiq}$\textquotedblright\ is one member of the set of super-operators in the right-hand side of equation (\ref{ppalt}).

Equivalence relation between quantum programs can be introduced based on their purely quantum denotational semantics.

\begin{defn}\label{eqpr} Let $P$ and $Q$ be two QGCL programs. Then: \begin{enumerate}\item We say that $P$ and $Q$ are equivalent and write $P\equiv Q$ if $$\llbracket P\rrbracket\otimes\mathcal{I}_{Q\setminus P}=\llbracket Q\rrbracket\otimes\mathcal{I}_{P\setminus Q},$$ where $\mathcal{I}_{Q\setminus P}$ is the identity super-operator in $\mathcal{H}_{qvar(Q)\setminus qvar(P)}$ and $\mathcal{I}_{P\setminus Q}$ the identity super-operator in $\mathcal{H}_{qvar(P)\setminus qvr(Q)}$.\item The \textquotedblleft coin-free\textquotedblright equivalence $P\equiv_{CF}Q$ holds if  \begin{equation}\label{cofree}tr_{\mathcal{H}_{cvar(P)\cup cvar(Q)}}(\llbracket P\rrbracket\otimes\mathcal{I}_{Q\setminus P})=tr_{\mathcal{H}_{cvar(P)\cup cvar(Q)}}(\llbracket Q\rrbracket\otimes\mathcal{I}_{P\setminus Q}).\end{equation}
\end{enumerate}
\end{defn}

If $qvar(P)=qvar(Q)$, then $P\equiv Q$ if and only if $\llbracket P\rrbracket=\llbracket Q\rrbracket$, and $P\equiv_{CF} Q$ if and only if $tr_{\mathcal{H}_{cvar(P)}}(\llbracket P\rrbracket)=tr_{\mathcal{H}_{cvar(P)}}\llbracket Q\rrbracket$. The symbol \textquotedblleft$tr$\textquotedblright\ in equation (\ref{deftr}) denotes partial trace, which is defined s follows: let $\mathcal{H}_1$ and $\mathcal{H}_2$ be two Hilbert spaces and let $\{|\varphi_i\rangle\}$ be an orthonormal basis of $\mathcal{H}_1$. Then for any density operator $\rho$ in $\mathcal{H}_1\otimes\mathcal{H}_2$, \begin{equation}\label{trout}tr_{\mathcal{H}_1}(\rho)=\sum_i\langle\varphi_i|\rho|\varphi_i\rangle\end{equation} is a density operator in $\mathcal{H}_2$. Furthermore, for any super-operator $\mathcal{E}$ on $\mathcal{H}_1\otimes\mathcal{H}_2$, $tr_{\mathcal{H}_1}(\mathcal{E})$ is a super-operator from $\mathcal{H}_1\otimes\mathcal{H}_2$ to $\mathcal{H}_2$ defined by $tr_{\mathcal{H}_1}(\mathcal{E})(\rho)=tr_{\mathcal{H}_1}(\mathcal{E}(\rho))$ for all density operators $\rho$ in $\mathcal{H}_1\otimes\mathcal{H}_2$. In a sense, \textquotedblleft coin\textquotedblright\ variables are only used to realize superposition of programs. The computational outcome of a program $P$ is stored in the \textquotedblleft principal\textquotedblright\ state space $\mathcal{H}_{qvar(P)\setminus cvar(P)}$. This is exactly the reason why we introduce the notion of \textquotedblleft coin-free\textquotedblright equivalence. Obviously, $P\equiv Q$ implies $P\equiv_{CF}Q$.

\subsection{Weakest Precondition Semantics}

The notions of Hoare triple for a quantum program and quantum weakest precondition were proposed by D'Hondt and Panangaden in \cite{DP06}. We now  recall their definitions from \cite{DP06}.

\begin{defn} Let $P$ be a program, and let $N_1$ and $N_2$ be (bounded) positive (Hermitian) operators in $\mathcal{H}_{qvar(P)}$. \begin{enumerate}
\item If $$tr(N_1\rho)\leq tr(N_2\llbracket P\rrbracket (\rho))$$ for all density operators $\rho$ in $\mathcal{H}_{qvar(P)}$, then $N_1$ is called a precondition of $N_2$ and $N_2$  a postcondition of $N_1$ with respect to $P$, and we write \begin{equation}\label{HoaT}\{N_1\}P\{N_2\}.\end{equation} Equation~(\ref{HoaT}) is called a (quantum) Hoare triple. \item $N_2$ is called the weakest precondition of $N_1$ with respect to $P$, written $N_2=wp.P.N_1$ if \begin{enumerate}\item $N_2$ is a precondition of $N_1$ with respect to $P$; and\item $N^\prime\sqsubseteq N_2$ whenever $N^\prime$ is a also precondition of $N_1$ with respect to $P$, where $\sqsubseteq$ stands for the L\"{o}wner order.
\end{enumerate}\end{enumerate}
\end{defn}

\begin{rem}In the original definition of quantum weakest precondition in \cite{DP06}, $N_1$ and $N_2$ are required to be so-called quantum predicates; i.e. Hermitian operators whose eigenvalues are in the unit interval $[0,1]$. However, this constraint is not essential, and thus it was removed in the above definition. Allowing $N_1$ and $N_2$ to be any (bounded) positive operators is indeed consistent with the treatment of probabilistic weakest preconditions in \cite{Ko81,MM05}, where a probabilistic precondition or postcondition was understood as the expectation of a random variable.\end{rem}

For each program $P$, $wp.P$ can be seen as the super-operator in $\mathcal{H}_{qvar(P)}$ defined as follows: for any positive operator $N$, $$(wp.P)(N)=wp.P.N$$ is given by clause (2) of the above definition, and $wp.P$ can be extended to the whole space of bounded operators in $\mathcal{H}_{qvar(P)}$ by linearity.

The weakest precondition semantics of QGCL programs can be derived from Proposition~\ref{qsem}, and they are given in the next proposition.

\begin{prop}\label{wpc}\begin{enumerate}\item $wp.\mathbf{abort}=0;$\item $wp.\mathbf{skip}=1;$\item $wp.(P_1;P_2)=wp.P_2;wp.P_1;$ \item $wp.U[\overline{q}]=U_{\overline{q}}^\dag\circ U_{\overline{q}};$\item $wp.\mathbf{measure}\ \left(\square m\cdot M[\overline{q}:x]=m\rightarrow P_m\right)\ \mathbf{end}=\sum_m\left[wp.P_m;(M_m^\dag\circ M_m)\right];$\item $wp.\mathbf{qif}\ [\overline{q}] \left(\square i\cdot |i\rangle\rightarrow P_i\right)\ \mathbf{fiq}\in \square_{i}\ |i\rangle\rightarrow wp.P_i.$
\end{enumerate}\end{prop}

Some cylindrical extensions of super-operators are used but unspecified in the above proposition because they can be recognised from the context. Again, \textquotedblleft$\in$" in the above clause (6) cannot be replaced by equality because the right-hand side of clause (6) is a set that may contain more than one super-operator.

We can define the refinement relation between quantum programs in terms of their weakest precondition semantics. To this end, we first generalize the L\"{o}wner order to the case of super-operators: for any two super-operators $\mathcal{E}$ and $\mathcal{F}$ in Hilbert space $\mathcal{H}$, $\mathcal{E}\sqsubseteq\mathcal{F}$ if and only if $\mathcal{E}(\rho)\sqsubseteq\mathcal{F}(\rho)$ for all density operators $\rho$ in $\mathcal{H}$.

\begin{defn}Let $P$ and $Q$ be two programs. Then we say that $P$ is refined by $Q$ and write $P\sqsubseteq Q$ if $$wp.P\otimes\mathcal{I}_{Q\setminus P}\sqsubseteq wp.Q\otimes\mathcal{I}_{P\setminus Q},$$ where $\mathcal{I}_{Q\setminus P}$ and $\mathcal{I}_{P\setminus Q}$ are the same as in Deinition~\ref{eqpr}.\end{defn}

It is easy to see that $P\sqsubseteq Q$ and $Q\sqsubseteq P$ implies $P\equiv Q$. Here, we are not going to further consider how can refinement technique be used in quantum programming, but leave it as a topic for future research.

\subsection{An Example}

To conclude this section, we present a simple example that helps us to understand the semantic notions introduced above.

\begin{exam} Let $q$ be a qubit variable and $x, y$ two classical variables. Consider the QGCL program
\begin{equation*}\begin{split}P\stackrel{\triangle}{=}\mathbf{qif}\ |0\rangle\rightarrow\ &H[q];\\
&\mathbf{measure}\ M^{(0)}[x\leftarrow q]=0\rightarrow X[q];\\
&\ \ \ \square\ \ \ \ \ \ \ \ \ \ \ \ \ \ \ \ \ \ \ \ \ \ \ \ \ \ \ \ \ \ \ \ \ \ \ \ \ \ 1\rightarrow Y[q]\\
&\mathbf{end}\\
\square\ |1\rangle\rightarrow\ &S[q];\\
&\mathbf{measure}\ M^{(1)}[x\leftarrow q]=0\rightarrow Y[q]\\
&\ \ \ \square\ \ \ \ \ \ \ \ \ \ \ \ \ \ \ \ \ \ \ \ \ \ \ \ \ \ \ \ \ \ \ \ \ \ \ \ \ \ 1\rightarrow Z[q]\\
&\mathbf{end};\\
&X[q];\\
&\mathbf{measure}\ M^{(0)}[y\leftarrow q]=0\rightarrow Z[q]\\
&\ \ \ \square\ \ \ \ \ \ \ \ \ \ \ \ \ \ \ \ \ \ \ \ \ \ \ \ \ \ \ \ \ \ \ \ \ \ \ \ \ \ 1\rightarrow X[q]\\
&\mathbf{end}\\
\mathbf{fiq}\ \ \ \ \ \ \ \ \ \ \ \ &
\end{split}\end{equation*}where $M^{(0)}, M^{(1)}$ are the measurements on a qubit in computational basis $|0\rangle, |1\rangle$ and basis $|\pm\rangle$, respectively (see Example \ref{gmeas}), $H$ is the Hadamard gate, $$X=\left(\begin{array}{cc} 0&1\\ 1& 0\end{array}\right),\ \ Y=\left(\begin{array}{cc} 0& -i\\ i& 0\end{array}\right),\ \ Z=\left(\begin{array}{cc} 1&0\\ 0& -1\end{array}\right)$$ are the Pauli matrices, and $$S=\left(\begin{array}{cc} 1&0\\ 0& i\end{array}\right)$$ is the phase gate. The program $P$ is a quantum alternation between two subprograms $P_0$ and $P_1$. The first subprogram $P_0$ is the Hadamard gate followed by the measurement in the computational basis, where whenever the outcome is $0$, then the gate $X$ follows; whenever the outcome is $1$, then the gate $Y$ follows. The second subprogram $P_1$ is the gate $S$ followd by the measurement in basis $|\pm\rangle$, the gate $X$, and the measurement in the computational basis.

We write $a$ for classical state $[x\leftarrow a]$ of program $P_0$ and $bc$ for classical state $[x\leftarrow b,y\leftarrow c]$ of program $P_1$ for any $a, c\in\{0,1\}$ and $b\in\{+,-\}$. Then the semi-classical semantic functions of $P_0$ and $P_1$ are given as follows: \begin{equation*}\begin{cases}\lceil P_0\rceil(0)=X\cdot |0\rangle\langle 0|\cdot H=\frac{1}{\sqrt{2}}\left(\begin{array}{cc}0 &0\\ 1&1\end{array}\right),\\ \lceil P_0\rceil(1)=Y\cdot |1\rangle\langle 1|\cdot H=\frac{i}{\sqrt{2}}\left(\begin{array}{cc}-1 &1\\ 0&0\end{array}\right),\end{cases}\end{equation*}
\begin{equation*}\begin{cases}\lceil P_1\rceil(+0)=Z\cdot |0\rangle\langle 0|\cdot X\cdot Y\cdot |+\rangle\langle +|\cdot S=\frac{1}{2}\left(\begin{array}{cc}i &-1\\ 0&0\end{array}\right),\\ \lceil P_1\rceil(+1)=X\cdot |1\rangle\langle 1|\cdot X\cdot Y\cdot |+\rangle\langle +|\cdot S=\frac{1}{2}\left(\begin{array}{cc}-i &1\\ 0&0\end{array}\right),\\\lceil P_1\rceil(-0)=Z\cdot |0\rangle\langle 0|\cdot X\cdot Z\cdot |-\rangle\langle -|\cdot S=\frac{1}{2}\left(\begin{array}{cc}1 &-i\\ 0&0\end{array}\right),\\ \lceil P_1\rceil(-1)=X\cdot |1\rangle\langle 1|\cdot X\cdot Z\cdot |-\rangle\langle -|\cdot S=\frac{1}{2}\left(\begin{array}{cc}1 &-i\\ 0&0\end{array}\right).\end{cases}\end{equation*} The semi-classical semantic function of $P$ is an operator-valued function in the state space of two qubits over classical states $\Delta (P)=\{a\oplus bc: a,c\in\{0,1\}\ {\rm and}\ b\in\{+,-\}\}.$ It follows from equation (\ref{coef0}) that  \begin{equation*}\begin{split}\lceil P\rceil (a\oplus bc)(|0\rangle |\varphi\rangle)&=\lambda_{1(bc)}|0\rangle (\lceil P_0\rceil (a)|\varphi\rangle),\\ \lceil P\rceil (a\oplus bc)(|1\rangle |\varphi\rangle)&=\lambda_{0a}|1\rangle (\lceil P_1\rceil (bc)|\varphi\rangle),\end{split}\end{equation*} where $\lambda_{0a}=\frac{1}{\sqrt{2}}$ and $\lambda_{1(bc)}=\frac{1}{2}$ for $a,c\in\{0,1\}$ and $b\in\{+,-\}$. Using $$\lceil P\rceil (a\oplus bc)=\sum_{i,j\in{0,1}}(\lceil P\rceil (a\oplus bc)|ij\rangle)\langle ij|,$$ we can compute:
\begin{equation*}\begin{split}\lceil P\rceil (0\oplus +0)&=\frac{1}{2\sqrt{2}}\left(\begin{array}{cccc}0 & 1&0&0\\ 0&1&0&0\\ 0&0&i&0\\ 0&0&-1&0\end{array}\right),\ \ \ \lceil P\rceil (0\oplus +1)=\frac{1}{2\sqrt{2}}\left(\begin{array}{cccc}0 & 1&0&0\\ 0&1&0&0\\ 0&0&-i&0\\ 0&0&1&0\end{array}\right),\\
\lceil P\rceil (0\oplus -0)&=\lceil P\rceil (0\oplus -1)=\frac{1}{2\sqrt{2}}\left(\begin{array}{cccc}0 & 1&0&0\\ 0&1&0&0\\ 0&0&1&0\\ 0&0&-i&0\end{array}\right),\\
\lceil P\rceil (1\oplus +0)&=\frac{1}{2\sqrt{2}}\left(\begin{array}{cccc}-1 & 0&0&0\\ 1&0&0&0\\ 0&0&i&0\\ 0&0&-1&0\end{array}\right),\ \ \
\lceil P\rceil (1\oplus +1)=\frac{1}{2\sqrt{2}}\left(\begin{array}{cccc}-1 & 0&0&0\\ 1&0&0&0\\ 0&0&-i&0\\ 0&0&1&0\end{array}\right),\\
\lceil P\rceil (1\oplus -0)&=\lceil P\rceil (1\oplus -1)=\frac{1}{2\sqrt{2}}\left(\begin{array}{cccc}1 & 0&0&0\\ 1&0&0&0\\ 0&0&1&0\\ 0&0&-i&0\end{array}\right).\end{split}\end{equation*} Then the purely quantum semantics of program $P$ is the super-operator:
$$\llbracket P\rrbracket =\sum_{a,c\in\{0,1\}\ {\rm and}\ b\in \{+,-\}} E_{abc}\circ E_{abc}^\dag,$$ and it follows from the proof of Proposition 3.3  in \cite{DP06} that the weakest precondition semantics of $P$ is the super-operator
$$wp.P =\sum_{a,c\in\{0,1\}\ {\rm and}\ b\in \{+,-\}} E_{abc}^\dag \circ E_{abc},$$ where $E_{abc}
=\lceil P\rceil (a\oplus bc)$.
\end{exam}

\section{Quantum Choice}\label{QChoice}

As discussed in Subsection~\ref{decin}, quantum alternation and choice are two ingredients in the realization of the quantum programming paradigm of superposition of programs. But only quantum alternation was introduced as a primitive program construct in the syntax of QGCL. Indeed, as already explained in Subsection~\ref{decin}, quantum choice may be easily defined as a derived program construct from quantum alternation.

\begin{defn}\label{cho-def}
Let $P$ be a program such that $\overline{q}=qvar(P)$, and let $P_i$ be programs for all $i$ . If $\{|i\rangle\}$ is an orthonormal basis of $\mathcal{H}_{\overline{q}}$, and $\overline{q}\cap \bigcup_{i} qvar(P_i)=\emptyset$, then the quantum choice of $P_i$'s according to $P$ along the basis $\{|i\rangle\}$ is defined as $$[P]\left(\bigoplus_{i}|i\rangle \rightarrow P_i\right)\stackrel{\triangle}{=}P; \mathbf{qif}\ [\overline{q}] \left(\square i\cdot |i\rangle \rightarrow P_i\right)\ \mathbf{fiq}.$$ In particular, if $n=2$, then the quantum choice will be abbreviated to $P_0\ _{P}\oplus P_1.$
\end{defn}

Since the quantum choice of $P_1,...,P_n$ is defined in terms of their quantum alternation, the semantics of the former can be directly derived from that of the latter.

\subsection{Quantum Implementation of Probabilistic Choice}

The relationship between probabilistic choice and quantum choice was briefly discussed at the end of Subsection~\ref{decin}. Now it is the right time to examine this relationship in a more precise way. To this end, we first expand the syntax and semantics of QGCL to include probabilistic choice \cite{MM05}.

\begin{defn}\label{prob-def}Let $P_i$ be a QGCL program for each $1\leq i\leq n$, and let $\{p_i\}_{i=1}^n$ be a sub-probability distribution; that is, $p_i> 0$ for each $1\leq i\leq n$ and $\sum_{i=1}^np_i\leq 1$. Then \begin{enumerate}\item The probabilistic choice of $P_1,...,P_n$ according to $\{p_i\}_{i=1}^n$ is
$$\sum_{i=1}^n P_i@p_i.$$ \item The quantum variables of the choice are: $$qvar\left(\sum_{i=1}^n P_i@p_i\right)=\bigcup_{i=1}^n qvar(P_i).$$ \item The purely quantum denotational semantics of the choice is:
\begin{equation}\label{ppcom}\left\llbracket \sum_{i=1}^n P_i@p_i\right\rrbracket =\sum_{i=1}^n p_i\cdot \llbracket P_i\rrbracket.\end{equation}\end{enumerate}
\end{defn}

The right-hand side of equation (\ref{ppcom}) is the probabilistic combination of super-operators $\llbracket P_i\rrbracket$ according to distribution $\{p_i\}$; that is, $$\left(\sum_{i=1}^n p_i\cdot \llbracket P_i\rrbracket\right)(\rho)=\sum_{i=1}^n p_i\cdot \llbracket P_i\rrbracket(\rho)$$ for all density operators $\rho$. It is obvious that $\sum_{i=1}^n p_i\cdot \llbracket P_i\rrbracket$ is a super-operator too.

A clear description about the relationship between probabilistic choice and quantum choice requires us to further expand the syntax and semantics of QGCL by introducing block command with local quantum variables.

\begin{defn}\label{loc-def} Let $P$ be a QGCL program, let $\overline{q}\subseteq qvar(P)$ be a sequence of quantum variables, and let $\rho$ be a density operator in $\mathcal{H}_{\overline{q}}$. Then \begin{enumerate}\item The block command defined by $P$ restricted to $\overline{q}=\rho$ is: \begin{equation}\label{blocm}\mathbf{begin\ local}\ \overline{q}:=\rho;P\ \mathbf{end}.\end{equation}\item The quantum variables of the block command are:
$$qvar\left(\mathbf{begin\ local}\ \overline{q}:=\rho;P\ \mathbf{end}\right)=qvar(P)\setminus\overline{q}.$$ \item The purely quantum denotational semantics of the block command is give as follows: \begin{equation}\label{deftr}\left\llbracket \mathbf{begin\ local}\ \overline{q}:=\rho;P\ \mathbf{end}\right\rrbracket (\sigma)=tr_{\mathcal{H}_{\overline{q}}}(\llbracket P\rrbracket (\sigma\otimes\rho))\end{equation} for any density operator $\sigma$ in $\mathcal{H}_{qvar(P)\setminus\overline{q}}$.
\end{enumerate}\end{defn}

The intuitive meaning of block command (\ref{blocm}) is that program $P$ is running in the environment where $\overline{q}$ are local variables and they are initialized in state $\rho$ before the execution of $P$. The symbol \textquotedblleft$tr$\textquotedblright\ in equation (\ref{blocm}) is partial trace defined by equation (\ref{trout}). So, after executing $P$, the auxiliary system denoted by the local variables $\overline{q}$ is discarded.

We present a simple example to illustrate the above two definitions.

\begin{exam} (Continuation of Example~\ref{gmeas}; Probabilistic mixture of measurements) It is often required in quantum cryptographic protocols like BB84 to randomly choose between the measurement $M^{(0)}$ on a qubit in the computational basis and the measurement $M^{(1)}$ in the basis $|\pm\rangle$. Here we consider a simplified version of random choice between $M^{(0)}$ and $M^{(1)}$. If we perform measurement $M^{(0)}$ on qubit $q$ in state $|\psi\rangle$ and discard the outcomes of measurement, then we get $$\rho_0=M_0^{(0)}|\psi\rangle\langle\psi|M_0^{(0)} +M_1^{(0)}|\psi\rangle\langle\psi|M_1^{(0)},$$ and if we perform measurement $M^{(1)}$ on  $|\psi\rangle$ and discard the outcomes, then we get $$\rho_1=M_+^{(1)}|\psi\rangle\langle\psi|M_+^{(1)} +M_-^{(1)}|\psi\rangle\langle\psi|M_-^{(1)}.$$
We now take the unitary matrix $$U=\left(\begin{array}{cc}\sqrt{p} & \sqrt{r}\\ \sqrt{r} & -\sqrt{p}\end{array}\right)$$ where $p,r\geq 0$ and $p+r=1$, and introduce a \textquotedblleft coin\textquotedblright\ qubit $q_C$. Let
\begin{equation*}\begin{split}P_i\stackrel{\triangle}{=}\mathbf{measure}\ M^{(i)}[x\leftarrow q]=\ &0\rightarrow \mathbf{skip}\\ \ \ \ \square\ \ \ \ \ \ \ \ \ \ \ \ \ \ \ \ \ \ \ \ \ \ \ \ \ \ \ \ \ \ \ \ \ \ \ &1\rightarrow \mathbf{skip}\\ \mathbf{end}\ \ \ \ \ \ \ \ \ \ \ \ \ \ \ \ \ \ \ \ \ \ \ \ \ \ \ \ \ \ \ \ \ & \end{split}\end{equation*} for $i=0,1$, and put quantum choice of $P_0$ and $P_1$ according the \textquotedblleft coin tossing operator\textquotedblright\ $U$ into a block with the \textquotedblleft coin\textquotedblright\ qubit $q_C$ as a local variable:
\begin{equation*}\begin{split}
P\stackrel{\triangle}{=}\ &\mathbf{begin\ local}\ q_C:=|0\rangle; P_0\ _{U[q_C]}\oplus P_1\ \mathbf{end}
\end{split}\end{equation*} Then for any $|\psi\rangle\in\mathcal{H}_{q}$, $i\in\{0,1\}$ and $j\in\{+,-\}$, we have: \begin{equation*}|\psi_{ij}\rangle\stackrel{\triangle}{=}M_{ij}(U|0\rangle|\psi\rangle)
=\sqrt{\frac{p}{2}}|0\rangle M_i^{(0)} |\psi\rangle+\sqrt{\frac{r}{2}} |1\rangle M_j^{(1)}|\psi\rangle,\end{equation*} \begin{equation*}\begin{split}
\llbracket P\rrbracket (|\psi\rangle\langle\psi|)&=tr_{\mathcal{H}_{q_C}}\left(\sum_{i\in\{0,1\}\ {\rm and}\ j\in\{+,-\}}|\psi_{ij}\rangle\langle\psi_{ij}|\right)\\
&=2 \sum_{i\in\{0,1\}}\frac{p}{2}M_i^{(0)}|\psi\rangle\langle\psi|M_i^{(0)} + 2\sum_{j\in\{+,-\}}\frac{r}{2}M_j^{(1)}|\psi\rangle\langle\psi|M_j^{(1)}
\\ &=p\rho_0 +r\rho_1.
\end{split}\end{equation*} So, program $P$ can be seen as a probabilistic mixture of measurements $M^{(0)}$ and $M^{(1)}$.
\end{exam}

Now we are ready to precisely characterize the relationship between probabilistic choice and quantum choice. Roughly speaking, if the \textquotedblleft coin" variables are treated as local variables, then a quantum choice degenerates to a probabilistic choice.

\begin{thm}\label{proim} Let $qvar(P)=\overline{q}$. Then we have:
\begin{equation}\label{ppim}\mathbf{begin\ local}\ \overline{q}:=\rho;[P]\left(\bigoplus_{i=1}^n|i\rangle\rightarrow P_i\right)\ \mathbf{end}\equiv \sum_{i=1}^nP_i@p_i\end{equation} where $p_i=\langle i|\llbracket P\rrbracket (\rho)|i\rangle$ for every $1\leq i\leq n$.
\end{thm}

The inverse of the above theorem is also true. For any probability distribution $\{p_i\}_{i=1}^n$, we can find an $n\times n$ unitary operator $U$ such that $p_i=|U_{i0}|^2$ $(1\leq i\leq n)$. So, it follows immediately from the above theorem that a probabilistic choice $\sum_{i=1}^nP_i@p_i$ can always be implemented by a quantum choice: $$\mathbf{begin\ local}\ \overline{q}:=|0\rangle;[U[\overline{q}]]\left(\bigoplus_{i=1}^n |i\rangle\rightarrow P_i\right)\ \mathbf{end}$$
where $\overline{q}$ is a family of new quantum variables with an $n-$dimensional state space. As said in Subsection~\ref{12}, probabilistic choice (\ref{proch}) can be thought of as a refinement of nondeterministic choice (\ref{dech}). Since for a given probability distribution $\{p_i\}$, there are more than one \textquotedblleft coin program\textquotedblright\ $P$ to implement the probabilistic choice $\sum_{i=1}^nP_i@p_i$ in equation (\ref{ppim}), a quantum choice can be further seen as a refinement of a probabilistic choice where a specific \textquotedblleft device\textquotedblright\ (quantum \textquotedblleft coin\textquotedblright) is explicitly given for generating the distribution $\{p_i\}$.

\section{Algebraic Laws}\label{Alaws}

In this section, we present a group of basic algebraic laws for quantum alternation and choice, which will be useful for verification, transformation and compilation of quantum programs. The laws given in the following theorem shows that quantum alternation is idempotent, commutative and associative and sequential composition is distributive over quantum alternation from the right.

\begin{thm}\label{law0} (Laws for Quantum Alternation)\begin{enumerate}
\item Idempotent Law: If $P_i\equiv P$ for all $i$, then $$\mathbf{qif}\ \left(\square i\cdot |i\rangle\rightarrow P_i\right)\ \mathbf{fiq}\equiv P.$$ \item Commutative Law: For any permutation $\tau$ of $\{1,...,n\}$, we have:
$$\mathbf{qif}\ [\overline{q}] \left(\square_{i=1}^n i\cdot |i\rangle\rightarrow P_{\tau(i)}\right)\ \mathbf{fiq}\equiv U_{\tau^{-1}}[\overline{q}];\mathbf{qif}\ [\overline{q}]\left(\square_{i=1}^n i\cdot |i\rangle\rightarrow P_i\right)\ \mathbf{fiq};U_\tau[\overline{q}],$$ where $\tau^{-1}$ is the inverse of $\tau$, i.e. $\tau^{-1}(i)=j$ if and only if $\tau(j)=i$ for $i,j\in\{1,...,n\}$, and $U_\tau$ (resp. $U_{\tau^{-1}}$) is the unitary operator permutating the basis $\{|i\rangle\}$ of $\mathcal{H}_{\overline{q}}$ with $\tau$ (resp. $\tau^{-1}$); that is, $U_\tau(|i\rangle)=|\tau(i)\rangle$ (resp. $U_{\tau^{-1}}(|i\rangle)=|\tau^{-1}(i)\rangle$) for every $1\leq i\leq n$.
\item Associative Law: $$\mathbf{qif}\ \left(\square i\cdot |i\rangle\rightarrow \mathbf{qif}\ \left(\square j_i\cdot |j_i\rangle\rightarrow P_{ij_i}\right)\ \mathbf{fiq}\right)\ \mathbf{fiq}\equiv \mathbf{qif}\ (\overline{\alpha})\left(\square i,j_i\cdot |i,j_i\rangle\rightarrow P_{ij_i}\right)\ \mathbf{fiq}$$ for some family $\overline{\alpha}$ of parameters, where the right-hand side is a parameterized quantum alternation defined in Appendix A.
    \item Distributive Law: If $\overline{q}\cap qvar(Q)=\emptyset$, then $$\mathbf{qif}\ [\overline{q}]\left(\square i\cdot |i\rangle\rightarrow P_i\right)\ \mathbf{fiq};Q\equiv_{CF}\mathbf{qif}\ (\overline{\alpha})[\overline{q}]\left(\square i\cdot |i\rangle\rightarrow (P_i;Q)\right) \mathbf{fiq}$$ for some family $\overline{\alpha}$ of parameters, where the right-hand side is a parameterized quantum alternation. In particular, if we further assume that $Q$ contains no measurements, then $$\mathbf{qif}\ [\overline{q}]\left(\square i\cdot |i\rangle\rightarrow P_i\right)\ \mathbf{fiq};Q\equiv\mathbf{qif}\ [\overline{q}]\left(\square i\cdot |i\rangle\rightarrow (P_i;Q)\right) \mathbf{fiq}.$$
\end{enumerate}\end{thm}

A quantum choice is defined as a \textquotedblleft coin" program followed by a quantum alternation. A natural question would be: is it possible to move the \textquotedblleft coin" program to the end of a quantum alternation? The following theorem positively answers this question under the condition that encapsulation in a block with local variables is allowed.

\begin{thm}\label{local} For any programs $P_i$ and unitary operator $U$, we have:\begin{equation}\label{local1}[U[\overline{q}]]\left(\bigoplus_{i=1}^n |i\rangle\rightarrow P_i\right)\equiv \mathbf{qif}\ (\square i\cdot U^\dag_{\overline{q}}|i\rangle\rightarrow P_i)\ \mathbf{fiq};U[\overline{q}].\end{equation} More generally, for any programs $P_i$ and $P$ with $\overline{q}=qvar(P)$, there are new quantum variables $\overline{r}$, a pure state $|\varphi_0\rangle\in\mathcal{H}_{\overline{r}}$, an orthonormal basis $\{|\psi_{ij}\rangle\}$ of $\mathcal{H}_{\overline{q}}\otimes\mathcal{H}_{\overline{r}}$, programs $Q_{ij}$, and a unitary operator $U$ in $\mathcal{H}_{\overline{q}}\otimes\mathcal{H}_{\overline{r}}$ such that
\begin{equation}\label{local2}\begin{split}[P]\left(\bigoplus_{i=1}^n |i\rangle\rightarrow P_i\right)\equiv\ &\mathbf{begin\ local}\ \overline{r}:=|\varphi_0\rangle;\\ &\ \ \ \ \ \ \ \ \ \ \ \ \ \ \mathbf{qif}\
\left(\square i,j\cdot |\psi_{ij}\rangle\rightarrow Q_{ij}\right)\ \mathbf{fiq};\\ &\ \ \ \ \ \ \ \ \ \ \ \ \ \ U[\overline{q},\overline{r}]\\ &\mathbf{end}.\end{split}\end{equation}\end{thm}

The next theorem shows that quantum choice is also idempotent, commutative and associative and sequential composition is distributive over quantum choice from the right.

\begin{thm}\label{law} (Laws for Quantum Choice)\begin{enumerate}
\item Idempotent Law: If $qvar(Q)=\overline{q}$, $tr\llbracket Q\rrbracket (\rho)=1$ and $P_i\equiv P$ for all $1\leq i\leq n$, then $$\mathbf{begin\ local}\ \overline{q}:=\rho;[Q]\left(\bigoplus_{i=1}^n  |i\rangle\rightarrow P_i\right)\ \mathbf{end}\equiv P.$$ \item Commutative Law: For any permutation $\tau$ of $\{1,...,n\}$, we have:
$$[P]\left(\bigoplus_{i=1}^n |i\rangle\rightarrow P_{\tau(i)}\right)\equiv [P;U_\tau[\overline{q}]]\left(\bigoplus_{i=1}^n|i\rangle\rightarrow P_i\right);U_{\tau^{-1}} [q],$$ where $qvar(P)=\overline{q}$, and $U_\tau$, $U_{\tau^{-1}}$ are the same as in Theorem~\ref{law0} (2).
\item Associative Law: Let $\Gamma=\{(i,j_i):1\leq i\leq m\ {\rm and}\ 1\leq j_i\leq n_i\}=\bigcup_{i=1}^m(\{i\}\times\{1,...,n_i\}),$ and $$R=[P]\left(\bigoplus_{i=1}^n |i\rangle\rightarrow Q_i\right).$$ Then
\begin{equation*}\begin{split}[P]\left(\bigoplus_{i=1}^m |i\rangle \rightarrow [Q_i]\left(\bigoplus_{j_i=1}^{n_i} |j_i\rangle \rightarrow R_{ij_i}\right)\right)\equiv [R(\overline{\alpha})]\left(\bigoplus_{(i,j_i)\in\Gamma} |i,j_i\rangle\rightarrow R_{ij_i}\right),
\end{split}\end{equation*}
for some family $\overline{\alpha}$ of parameters, where the right-hand side is a parameterized quantum choice defined in Appendix A.
\item Distributive Law: If $qvar(P)\cap qvar(Q)=\emptyset$, then $$[P]\left(\bigoplus_{i=1}^n |i\rangle\rightarrow P_i\right);Q\equiv_{CF} [P(\overline{\alpha})]\left(\bigoplus_{i=1}^n |i\rangle\rightarrow (P_i;Q)\right)$$ for some family $\overline{\alpha}$ of parameters, where the right-hand side is a parameterized quantum choice. In particular, if we further assume that $Q$ contains no measurements, then $$[P]\left(\bigoplus_{i=1}^n |i\rangle\rightarrow P_i\right);Q\equiv [P]\left(\bigoplus_{i=1}^n |i\rangle\rightarrow (P_i;Q)\right).$$
\end{enumerate}\end{thm}

\section{Illustrative Examples}\label{IExam}

The design of the language QGCL, in particular the definition of quantum alternation and choice, was inspired by the construction of some simplest quantum walks. A large number of variants and generalizations of quantum walks have been introduced in the last decade. Quantum walks have been widely used in the development of quantum algorithms including quantum simulation. It was proved that quantum walks are indeed universal for quantum computation~\cite{Chi09,Lo10}. Furthermore, experimental implementations of quantum walks have also been conducting in the laboratories over the world. Various extended quantum walks in the literature can be conveniently written as QGCL programs with quantum alternation and choice. Here, we present several simple examples of quantum walks to further show the expressive power of QGCL.

\begin{exam}\label{wexam1} One of the simplest random walks is the one-dimensional walk where a walker moves to the left with probability $\frac{1}{2}$ and moves to the right with the same probability. The Hadamard walk considered in \cite{Am01} is a quantum generalization of this random walk. Let $p, c$ be the quantum variables for position and coin, respectively. The type of variable $p$ is the infinite-dimensional Hilbert space $$\mathcal{H}_p=span \{|n\rangle:n\in\mathbb{Z}\ (\rm integers)\}=\{\sum_{n=-\infty}^{\infty}\alpha_n |n\rangle: \sum_{n=-\infty}^{\infty}|\alpha_n|^2<\infty\},$$ and the type of $c$ is the $2-$dimensional Hilbert space
$\mathcal{H}_c=span\{|L\rangle,|R\rangle\}$, where $L, R$ stand for Left and Right, respectively. So, the state space of a walker on a line is $\mathcal{H}=\mathcal{H}_c\otimes\mathcal{H}_p$. We
write $I_{\mathcal{H}_p}$ for the identity operator in $\mathcal{H}_p$. Let $H$ be the $2\times 2$ Hadamard matrix (see equation (\ref{HDG})), and let $T_L, T_R$ be left- and right-translation, respectively; that is, $$T_L|n\rangle=|n-1\rangle,\ \ \ \ \ \ \ \ \ \ \ \ \ \ T_R|n\rangle=|n+1\rangle$$ for every $n\in \mathbb{Z}$.
Then a single step of the Hadamard walk can be described by the unitary operator \begin{equation}\label{stpop}W=(|L\rangle\langle L|\otimes T_L +|R\rangle\langle R|\otimes T_R)(H\otimes I_{\mathcal{H}_p}).\end{equation} It can also be written as the QGCL program: $$T_L[p] _{H[c]}\oplus T_R[p].$$ This program is the quantum choice of the left-translation $T_L$ and the right-translation $T_R$ according to the \textquotedblleft coin\textquotedblright\ program $H[c]$.
The Hadamard walk continuously runs this programs. The following are several variants of this walk considered in the recent physics literature.

\begin{enumerate}\item A simple variant of the above Hadamard walk is the unidirectional quantum walk examined in \cite{Mon13}, where the walk either moves to the right or stays in the previous position. So, the left-translation $T_L$ should be replaced by the program $\mathbf{skip}$ whose semantics is the identity operator $I_{\mathcal{H}_p}$, and a single step of the new quantum walk can be written as the QGCL program: $$\mathbf{skip}  _{H[c]}\oplus T_R[p].$$ It is a quantum choice of $\mathbf{skip}$ and the right-translation $T_R$.

\item A feature of the original one-dimensional quantum walk and its unidirectional variant is that the coin operator $H$ is independent of the position and time. A new kind of quantum walk was employed in \cite{KW13} to implement quantum measurement. The coin tossing operator of this walk depends on both position $n$ and time $t$:
$$C(n,t)=\frac{1}{\sqrt{2}}\left(\begin{array}{cc}c(n,t) & s(n,t)\\ s^\ast (n,t) &-e^{i\theta}c(n,t)\end{array}\right).$$ Then for a given time $t$, step $t$ of the walk can be written as the QGCL program:
\begin{equation*}\begin{split}W_t\stackrel{\triangle}{=}\ &\mathbf{qif}\ [p](\square n\cdot |n\rangle\rightarrow C(n,t)[c])\ \mathbf{fiq};\\ &\mathbf{qif}\ [c](|L\rangle\rightarrow T_L[p])\square (|R\rangle\rightarrow T_R[p])\ \mathbf{fiq}.
\end{split}\end{equation*} The program $W_t$ is a sequential composition of two quantum alternations. Since $W_t$ may be different for different time points $t$, the first $T$ steps can be written as the program: $$W_1;W_2;...;W_T.$$
\item Another simple generalization of the original and unidirectional one-dimensional quantum walk is the quantum walk with three coin states considered in~\cite{Inu05}. The coin space of this walk is a $3-$dimensional Hilbert space $\mathcal{H}_c=span\{|L\rangle, |0\rangle, |R\rangle\}$, where $L$ and $R$ are used to indicate moving to the left and to the right, respectively, as before, but $0$ means staying at the previous position. The \textquotedblleft coin tossing\textquotedblright\ operator is the unitary $$U=\frac{1}{3}\left(\begin{array}{ccc}-1 & 2 & 2\\ 2 & -1 & 2\\ 2& 2& -1\end{array}\right).$$ Then a single step of the walk can be written as the QGCL program: $$[U[c]]\left(|L\rangle\rightarrow T_L[p]\oplus |0\rangle\rightarrow \mathbf{skip}\oplus |R\rangle\rightarrow T_R[p]\right).$$
    This is the quantum choice of $\mathbf{skip}$, the left- and right-translations according to the \textquotedblleft coin\textquotedblright\ program $U[c]$.
    \end{enumerate}\end{exam}

    The quantum walks in the above example have only a single walker as well as a single \textquotedblleft coin\textquotedblright. In the following two examples, we consider some more complicated quantum walks in which multiple walkers participate and multiple \textquotedblleft coins\textquotedblright\ are equipped to control the walkers.

\begin{exam}A one-dimensional quantum walk driven by multiple coins was defined in~\cite{BCA03}. In this walk, there is still a single walker, but it is controlled by $M$ different \textquotedblleft coins\textquotedblright. Each of these \textquotedblleft coins\textquotedblright\ has its own state space, but the \textquotedblleft coin tossing\textquotedblright\ operator for all of them are the same, namely the $2\times 2$ Hadamard matrix. Now let variable $p$, space $\mathcal{H}_p, \mathcal{H}_c$ and operators $T_L, T_R, H$ are the same as in Example~\ref{wexam1}, and let $c_1,...,c_M$ be the quantum variables for the $M$ coins. Then the state space of the walk is $$\mathcal{H}=\mathcal{H}_p\otimes\bigotimes_{m=1}^M\mathcal{H}_{c_m},$$ where $\mathcal{H}_{c_m}=\mathcal{H}_c$ for all $1\leq m\leq M$.
 We write $$W_m=(T_L[p] _{H[c_1]}\oplus T_R[p]);...;(T_L[p] _{H[c_m]}\oplus T_R[p])$$ for $1\leq m\leq M$. If we cycle among the $M$ coins, starting from the coin $c_1$, then the first $T$ steps of the walk can be written in the language QGCL as follows: $$W_M;...;W_M;W_r$$ where $W_M$ is iterated for $d=\lfloor T/M\rfloor$ times, and $r=T-Md$ is the remainder of $T$ divided by $M$. This program is a sequential compositions of $T$ quantum choices of the left- and right translations controlled different \textquotedblleft coins\textquotedblright.\end{exam}

\begin{exam}\label{q2w} A quantum walk consisting of two walkers on a line sharing coins was introduced in~\cite{XS12}. In this walk, the two walkers have different state spaces, and each of the two walkers has its own \textquotedblleft coin\textquotedblright. So, the state Hilbert space of the whole quantum walk is $\mathcal{H}_p\otimes\mathcal{H}_p\otimes\mathcal{H}_c\otimes\mathcal{H}_c$. If the two walkers are completely independent, then the step operator of this walk is $W\otimes W$, where $W$ is defined by equation (\ref{stpop}). But more interesting is the case where a two-qubit unitary operator $U$ is introduced to entangle the two coins. This case can be thought of as that the two walkers are sharing coins. A step of this quantum walk can be written as a QGCL program as follows:
$$U[c_1,c_2];(T_L[q_1] _{H[c_1]}\oplus T_R[q_1]);(T_L[q_2] _{H[c_2]}\oplus T_R[q_2])$$ where $q_1, q_2$ are the position variables and $c_1,c_2$ the coin variables of the two walkers, respectively.
\end{exam}

\section{Conclusions}

In this paper, we introduce the notions of quantum alternation and choice by employing the idea of \textquotedblleft coin\textquotedblright\ systems used in the construction of quantum walks. They are quantum counterparts of the popular program construct of alternation, case statement or switch statement in classical programming languages and probabilistic choice in probabilistic programming languages. Based on them, a new quantum programming language, called QGCL, is defined. This language can be seen as a quantum generalization of Dijkstra's language GCL of guarded commands and Morgan et al.'s probabilistic programming language pGCL. It is also an extension of Sanders and Zuliani's quantum programming language qGCL. A salient feature of QGCL that all the previous quantum programming languages do not enjoy is that it can fully support a novel quantum programming paradigm - superposition of programs - which has been implicitly but widely used in the design of quantum walk-based algorithms. We believe that from the programming language point of view, the paradigm of superposition of programs will be a significant step to further exploit the power of quantum computing. This paper presents the denotational and weakest precondition semantics of the language QGCL, and establishes a group of basic algebraic laws that are useful in verification, transformation and compilation of QGCL programs.

We have developed a preliminary theory of quantum programming with quantum alternation and choice, but also leave a series of problems unsolved. Here, we list some of them for the future studies:

\begin{itemize}
\item The recursive construct of iteration (or while loop) can be conveniently defined in terms of alternation in classical programming languages. A kind of while loop for quantum programming was considered in \cite{Se04,YF10} based on the classical alternation (\ref{maltn}) of quantum programs, and it can be appropriately called \textit{classical controlled quantum loop}. How can we define \textit{quantum controlled loop} - loop based on quantum alternation introduced in this paper?
One of the key design ideas of almost all existing quantum programming languages can be summarised by the influential slogan \textquotedblleft quantum data, classical control\textquotedblright\ proposed by Selinger \cite{Se04}, meaning that the control flow of a quantum program is still classical, but the program operates on quantum data. An exception is Altenkirch and Grattage's functional language QML \cite{AG05}, where \textquotedblleft quantum control\textquotedblright\ flow was introduced. It seems that quantum alternation and choice together with quantum controlled loop will provide a much more general structure of control flows for quantum programming.

\item A quantum Floyd-Hoare logic was developed in~\cite{Ch06,Ka09,Y11} for quantum programs with only classical control flows. So, a further interesting problem  would be to extend this logic so that it can also be used to reasoning about programs with quantum control flows.

\item Of course, another important problem for further research is the implementation of the new quantum programming language QGCL. It is interesting to notice that recently physicists \cite{Zh11,Ar13} started to research on the physical implementation of a kind of control of quantum operations, which is similar to the guarded composition of two super-operators considered in Section \ref{sec-comp}.
\end{itemize}

\bibliographystyle{acmsmall}

\newpage

\appendix

\section{Choice of the Coefficients in Guarded Compositions of Quantum Operations}

The coefficients in the right-hand side of the defining equation (\ref{coef0}) of guarded composition of operator-valued function are chosen in a very special way with a physical interpretation in terms of conditional probability. This Appendix shows that other choices of these coefficients are possible. Let's first consider the guarded composition $$U\stackrel{\triangle}{=}\square_{k=1}^n |k\rangle\rightarrow U_k$$ of unitary operators $U_k$ $(1\leq k\leq n)$ in a Hilbert space $\mathcal{H}$ along an orthonormal basis $\{|k\rangle\}$ of a \textquotedblleft coin\textquotedblright\ Hilbert space $\mathcal{H}_C$. If for each $1\leq k\leq n$, we add a relative phase $\theta_k$ into the defining equation (\ref{defcu}) of $U$:
\begin{equation}\label{defcu1}U(|k\rangle |\psi\rangle)=e^{i\theta_k}|k\rangle(U_k|\psi\rangle)\end{equation} for all $|\psi\rangle\in\mathcal{H}$, then equation (\ref{eqUU}) is changed to \begin{equation}\label{eqUU1}U\left(\sum_{k,j}\alpha_{kj}|k\rangle|\psi_j\rangle\right)=\sum_{k,j}\alpha_{kj}e^{i\theta_k}|k\rangle(U_k|\psi_j\rangle).\end{equation}
It is easy to see that the new operator $U$ defined by equation (\ref{defcu1}) or (\ref{eqUU1}) is still unitary.

The idea of adding relative phases also applies to the guarded composition of operator-valued functions. Consider $$F\stackrel{\triangle}{=}\square_{k=1}^n|k\rangle\rightarrow F_k$$ where $\{|k\rangle\}$ is an orthonormal basis of $\mathcal{H}_C$, and $F_k$ is an operator-valued function in $\mathcal{H}$ over $\Delta_k$ for every $1\leq k\leq n$. We arbitrarily choose a sequence $\theta_1,...,\theta_n$ of real numbers and change the defining equation (\ref{coef0}) of $F$ to
\begin{equation}\label{coefnew}F(\oplus_{k=1}^n\delta_k)|\Psi\rangle=\sum_{k=1}^n e^{i\theta_k}\left(\prod_{l\neq k}\lambda_{l\delta_l}\right)|k\rangle(F_k(\delta_k)|\psi_k\rangle\end{equation} for any $|\Psi\rangle=\sum_{k=1}^n|k\rangle|\psi_k\rangle\in\mathcal{H}_C\otimes\mathcal{H}$, where $\lambda_{l\delta_l}$'s are the same as in Definition \ref{qgu}. Then it is clear that $F$ defined by equation (\ref{coefnew}) is still an operator-valued function. Indeed, this conclusion is true for a much more general guarded composition of operator-valued functions. Let $F_k$ be an operator-valued function in $\mathcal{H}$ over $\Delta_k$ for each $1\leq k\leq n$, and let \begin{equation}\label{parra}\overline{\alpha}=\left\{\alpha^{(k)}_{\delta_1,...,\delta_{k-1},\delta_{k+1},...,\delta_n}:1\leq k\leq n\ {\rm and}\ \delta_l\in\Delta_l\ {\rm for}\ l=1,...,k-1,k+1,...,n\right\}\end{equation} be a family of complex numbers satisfying the normalization condition: \begin{equation}\label{normc}\sum_{\delta_1\in\Delta_1,...,\delta_{k-1}\in\Delta_{k-1},\delta_{k+1}\in\Delta_{k+1},...,\delta_n\in\Delta_n}\left|
\alpha^{(k)}_{\delta_1,...,\delta_{k-1},\delta_{k+1},...,\delta_n}\right|^2=1\end{equation} for every $1\leq k\leq n$. Then we can define the $\overline{\alpha}-$guarded composition
$$F\stackrel{\triangle}{=}\left(\overline{\alpha}\right)\left(\square_{k=1}^n |i\rangle\rightarrow F_k\right)$$
of $F_k$ $(1\leq k\leq n)$ along an orthonormal basis $\{|k\rangle\}$ of $\mathcal{H}_C$ by \begin{equation}\label{gengu}F\left(\oplus_{k=1}^n\delta_k\right)\left(\sum_{k=1}^n |k\rangle|\psi_k\rangle\right)=\sum_{k=1}^n\alpha^{(k)}_{\delta_1,...,\delta_{k-1},\delta_{k+1},...,\delta_n} |k\rangle\left(F_k(\delta_k)|\psi_k\rangle\right)\end{equation} for any $|\psi_1\rangle,...,|\psi_n\rangle\in\mathcal{H}$ and for any $\delta_k\in\Delta_k$ $(1\leq k\leq n)$. Note that coefficient $\alpha^{(k)}_{\delta_1,...,\delta_{k-1},\delta_{k+1},...,\delta_n}$ does not contain parameter $\delta_k$. This independence together with condition (~\ref{normc}) guarantees that the $\overline{\alpha}-$guarded composition is an operator-valued function, as can be seen from the proof of Lemma~\ref{gulem} presented in Appendix \ref{C1}.

\begin{exam}\begin{enumerate}\item Definition \ref{qgu} is a special case of $\overline{\alpha}-$guarded composition because if for any $1\leq i\leq n$ and $\delta_k\in\Delta_k$ $(k=1,...,i-1,i+1,...,n)$, we set $$\alpha^i_{\delta_1,...,\delta_{i-1},\delta_{i+1},...,\delta_n}=\prod_{k\neq i}\lambda_{k\delta_k},$$ where $\lambda_{k\delta_k}$'s are given by equation (\ref{coef1}), then equation (\ref{gengu}) degenerates to (\ref{coef0}). \item Another possible choice of $\overline{\alpha}$ is $$\alpha^i_{\delta_1,...,\delta_{i-1},\delta_{i+1},...,\delta_n}=\frac{1}{\sqrt{\prod_{k\neq i}|\Delta_k|}}$$ for all $1\leq i\leq n$ and $\delta_k\in\Delta_k$ $(k=1,...,i-1,i+1,...,n)$. Obviously, for this family $\overline{\alpha}$ of coefficients, the $\overline{\alpha}-$guarded composition cannot be obtained by modifying Definition \ref{qgu} with relative phases.\end{enumerate}\end{exam}

Now we are able to define parameterized quantum alternation and choice, which are needed in the presentation of some algebraic laws in Section~\ref{Alaws}.

\begin{defn}\begin{enumerate}\item Let $\overline{q}$, $\{|i\rangle\}$ and $\{P_i\}$ be as in Definition \ref{syn-def} (4). Furthermore, let the classical states $\Delta(P_i)=\Delta_i$ for every $i$, and let $\overline{\alpha}$ be a family of parameters satisfying condition (\ref{normc}), as in equation (\ref{parra}). Then the $\overline{\alpha}-$quantum alternation of $P_1,...,P_n$ guarded by basis states $|i\rangle$'s is  \begin{equation}\label{ddqa1}P\stackrel{\triangle}{=}\mathbf{qif}\ (\overline{\alpha})[\overline{q}] \left(\square i\cdot \ |i\rangle\rightarrow P_i\right)\ \mathbf{fiq}\end{equation} and its semi-classical semantics is $$\lceil P\rceil =(\overline{\alpha})(\square_{i=1}^n |i\rangle\rightarrow\lceil P_i\rceil).$$
\item Let $P$, $\{|i\rangle\}$ and $P_i$'s be as in Definition \ref{cho-def}, and let $\overline{\alpha}$ be as above. Then the $\overline{\alpha}-$quantum choice of $P_i$'s according to $P$ along the basis $\{|i\rangle\}$ is defined as $$[P(\overline{\alpha})]\left(\bigoplus_{i}|i\rangle \rightarrow P_i\right)\stackrel{\triangle}{=}P; \mathbf{qif}\ (\overline{\alpha})[\overline{q}] \left(\square i\cdot |i\rangle \rightarrow P_i\right)\ \mathbf{end}.$$
\end{enumerate}
\end{defn}

The symbol $[\overline{q}]$ in quantum alternation (\ref{ddqa1}) can be dropped whenever quantum variables $\overline{q}$ can be recognized from the context. At the first glance, it seems unreasonable that the parameters $\overline{\alpha}$ in the syntax (\ref{ddqa1}) of $\overline{\alpha}-$quantum alternation are indexed by the classical states of $P_i$. But this is not problematic at all because the classical states of $P_i$ are completely determined by the syntax of $P_i$. The purely quantum denotational semantics of the $\overline{\alpha}-$quantum alternation can be obtained from its semi-classical semantics according to Definition \ref{den-def}, and the semantics of $\overline{\alpha}-$quantum choice can be derived from the semantics of $\overline{\alpha}-$quantum alternation.

\section{Quantum Alternation Guarded by Subspaces}\label{Orspace}

A major difference between alternation (\ref{altn1}) of classical programs and quantum alternation (\ref{ddqa}) can be revealed by a comparison between their guards: the guards $G_i$ in the former are propositions about the program variables, whereas the guards $|i\rangle$ in the latter are basis states of the \textquotedblleft coin\textquotedblright\ space $\mathcal{H}_C$. However, this difference is not as big as we imagine at the first glance. In the Birkhoff-von Neumann quantum logic \cite{BvN36}, a proposition about a quantum system is expressed by a closed subspace of the state Hilbert space of the system. This observation leads us to a way to define quantum alternation guarded by propositions about the \textquotedblleft coin\textquotedblright\ system instead of basis states of the \textquotedblleft coin\textquotedblright\ space.

\begin{defn} Let $\overline{q}$ be a sequence of quantum variables and $\{P_i\}$ be a family of programs such that $$\overline{q}\cap\left(\bigcup_{i}qVar(P_i)\right)=\emptyset.$$
Suppose that $\{X_i\}$ is a family of propositions about the \textquotedblleft coin\textquotedblright\ system $\overline{q}$, i.e. closed subspaces of the \textquotedblleft coin\textquotedblright\ space $\mathcal{H}_{\overline{q}}$, satisfying the following two conditions:\begin{enumerate}\item $X_i$'s are pairwise orthogonal, i.e. $X_{i_1}\bot X_{i_2}$ provided $i_1\neq i_2$; \item $\bigoplus_i X_i\stackrel{\triangle}{=}span \left(\bigcup_i X_i\right)=\mathcal{H}_{\overline{q}}$.\end{enumerate} Then \begin{enumerate}\item The quantum alternation of $P_i$'s guarded by subspaces $X_i$'s: \begin{equation}\label{gprop}P\stackrel{\triangle}{=}\mathbf{qif}\ [\overline{q}] \left(\square i\cdot X_i\rightarrow P_i\right)\ \mathbf{fiq}\end{equation} is a program.
\item The quantum variables of the alternation are: $$qVar(P)=\overline{q}\cup\left(\bigcup_{i}qVar(P_i)\right).$$ \item The purely quantum denotational semantics of the alternation is:
\begin{equation}\label{gprop1}\begin{split}\llbracket P\rrbracket=\{&\llbracket\mathbf{qif}\ [\overline{q}] \left(\square i,j_i\cdot |\varphi_{ij_i}\rangle
\rightarrow P_{ij_i}\right)\ \mathbf{fiq}\rrbracket:\{|\varphi_{ij_i}\rangle\}\ {\rm is\ an\ orthonormal}\\ &\ \ \ \ \ \ \ \ \ \ \ \ \ \ \ \ \ \ \ \ \ \ \ \ \ \ \ \ \ \ \ \ \ {\rm basis\ of}\ X_i\ {\rm for\ each}\ i,\ {\rm and}\ P_{ij_i}=P_i\ {\rm for\ every}\ i,j_i\}.\end{split}\end{equation}
\end{enumerate}\end{defn}

For simplicity, the variables $\overline{q}$ in quantum alternation (\ref{gprop}) can be dropped if they can be recognized from or irrelevant in the context. It is clear that the union $\bigcup_i \{|\varphi_{ij_i}\rangle\}$ of the bases of subspaces $X_i$'s in equation (\ref{gprop1}) is an orthonormal basis of the whole \textquotedblleft coin\textquotedblright\ space $\mathcal{H}_C$. Note that the purely quantum semantics of alternation (\ref{gprop}) guarded by subspaces is a set of super-operators rather than a single super-operator. So, alternation (\ref{gprop}) is a nondeterministic program, and its nondeterminism comes from different choices of the bases of guard subspaces. Furthermore, an alternation guarded by basis states of these subspaces is a refinement of alternation (\ref{gprop}).

The notion of program equivalence in Definition~\ref{eqpr} can be easily generalized to the case of nondeterministic programs provided we make the following conventions: \begin{itemize}\item If $\Omega$ is a set of super-operators and $\mathcal{F}$ a super-operator, then $$\Omega\otimes\mathcal{F}=\{\mathcal{E}\otimes\mathcal{F}:\mathcal{E}\in\Omega\};$$ \item We identify a single super-operator with the set containing only this super-operator.\end{itemize} Some basic properties of quantum alternation guarded by subspaces are given in the following:
\begin{prop}\begin{enumerate}\item If $P_i$ does not contain any measurement for all $i$, then for any orthonormal basis $\{|\varphi_{ij_i}\rangle\}$ of $X_i$ $(1\leq i\leq n)$, we have:
$$\mathbf{qif}\ (\square i\cdot X_i\rightarrow P_i)\ \mathbf{fiq}\equiv \mathbf{qif}\ (\square i,j_i\cdot |\varphi_{ij_i}\rangle \rightarrow P_{ij_i})\ \mathbf{fiq}$$ where $P_{ij_i}=P_i$ for every $i,j_i$. In particular, if $U_i$ is an unitary operator in $\mathcal{H}_{\overline{q}}$ for all $i$, then $$\mathbf{qif}\ [\overline{q}_C] (\square i\cdot X_i\rightarrow U_i[\overline{q}])\ \mathbf{fiq}\equiv U[\overline{q}_C, \overline{q}]$$ where $U=\sum_i (I_{X_i}\otimes U_i)$ is an unitary operator in $\mathcal{H}_{\overline{q}_C\cup \overline{q}}$.
\item Let $U$ be a unitary operator in $\mathcal{H}_{\overline{q}}$. If for every $i$, $X_i$ is an invariant subspace of $U$, i.e. $UX_i=\{U|\psi\rangle:|\psi\rangle\in X_i\}\subseteq X_i$, then $$U[\overline{q}];\mathbf{qif}\ [\overline{q}] \left(\square i\cdot X_i\rightarrow P_i\right)\ \mathbf{fiq}; U^\dag[\overline{q}]\equiv \mathbf{qif}\ [\overline{q}] \left(\square i\cdot X_i\rightarrow P_i\right)\ \mathbf{fiq}.$$\end{enumerate}\end{prop}

\section{Proofs of Lemmas, Propositions and Theorems}

\subsection{Proof of Lemma~\ref{gulem}}\label{C1}

Clause (2) can be proved by a routine calculation, which is omitted here. To prove clause (1), we write: $$\overline{F}\stackrel{\triangle}{=}\sum_{\delta_1\in\Delta_1,...,\delta_n\in\Delta_n}F(\oplus_{i=1}^n \delta_i)^\dag\cdot F(\oplus_{i=1}^n\delta_i).$$ Our purpose is to show that $\overline{F}\sqsubseteq I_{\mathcal{H}_C\otimes\mathcal{H}}$, and $\overline{F}= I_{\mathcal{H}_C\otimes\mathcal{H}}$ whenever all $F_i$'s are full. To do this, we start with an auxiliary equality. For any $|\Phi\rangle, |\Psi\rangle\in\mathcal{H}_C\otimes\mathcal{H}$, we can write: \begin{equation*}|\Phi\rangle=\sum_{i=1}^n|i\rangle|\varphi_i\rangle,\ \ \ \ \ \ \ \  |\Psi\rangle=\sum_{i=1}^n|i\rangle|\psi_i\rangle\end{equation*} where $|\varphi_i\rangle, |\psi_i\rangle\in\mathcal{H}$ for each $1\leq i\leq n$. Then we have: \begin{equation}\label{gud-pr}\begin{split}\langle\Phi|\overline{F}|\Psi\rangle&=\sum_{\delta_1,...,\delta_n}\langle\Phi| F(\oplus_{i=1}^n\delta_i)^\dag\cdot F(\oplus_{i=1}^n\delta_i)|\Psi\rangle\\
&=\sum_{\delta_1,...,\delta_n}\sum_{i,i^\prime =1}^n\left(\prod_{k\neq i}\lambda^\ast_{k\delta_k}\right)\left(\prod_{k\neq i^\prime}\lambda_{k\delta_k}\right)\langle i|i^\prime\rangle\langle\varphi_i|
F_{i}(\delta_i)^\dag F_{i^\prime}(\delta_{i^\prime})|\psi_{i^\prime}\rangle\\ &=\sum_{\delta_1,...,\delta_n}\sum_{i=1}^n\left(\prod_{k\neq i}|\lambda_{k\delta_k}|^2\right) \langle\varphi_i|F_{i}(\delta_i)^\dag F_{i}(\delta_{i})|\psi_{i}\rangle\\
&=\sum_{i=1}^n\left[\sum_{\delta_1,...,\delta_{i-1},\delta_{i+1},...,\delta_n}\left(\prod_{k\neq i}|\lambda_{k\delta_k}|^2\right)\cdot\sum_{\delta_i} \langle\varphi_i|F_{i}(\delta_i)^\dag F_{i}(\delta_{i})|\psi_{i}\rangle\right]\\
&=\sum_{i=1}^n \sum_{\delta_i} \langle\varphi_i|F_{i}(\delta_i)^\dag F_{i}(\delta_{i})|\psi_{i}\rangle\\
&=\sum_{i=1}^n\langle\varphi_i|\sum_{\delta_i}F_{i}(\delta_i)^\dag F_{i}(\delta_{i})|\psi_{i}\rangle
\end{split}\end{equation} because for each $k$, we have: $$\sum_{\delta_k}\left|\lambda_{k\delta_k}\right|^2=1,$$ and thus \begin{equation}\label{coeff}\sum_{\delta_1,...,\delta_{i-1},\delta_{i+1},...,\delta_n}\left(\prod_{k\neq i}|\lambda_{k\delta_k}|^2\right)
=\prod_{k\neq i}\left(\sum_{\delta_k}|\lambda_{k\delta_k}|^2\right)=1.\end{equation}

Now we are ready to prove our conclusions by using equation (\ref{gud-pr}).

(1) We first prove that $\overline{F}\sqsubseteq I_{\mathcal{H}_C\otimes\mathcal{H}}$, i.e. $F$ is an operator-valued function in $\mathcal{H}_C\otimes\mathcal{H}$ over $\bigoplus_{i=1}^n\Delta_n$. It suffices to show that $\langle\Phi|\overline{F}|\Phi\rangle\leq\langle\Phi|\Phi\rangle$ for each $|\Phi\rangle\in\mathcal{H}_C\otimes\mathcal{H}$. In fact, for each $1\leq i\leq n$, since $F_i$ is an operator-valued function, we have: $$\sum_{\delta_i}F_{i}(\delta_i)^\dag F_{i}(\delta_i)\sqsubseteq I_\mathcal{H}.$$ Therefore, it holds that $$\langle\varphi_i|\sum_{\delta_i}F_{i}(\delta_i)^\dag F_{i}(\delta_i)|\varphi_i\rangle\leq\langle\varphi_i|\varphi_i\rangle.$$ Then it follows immediately from equation~(\ref{gud-pr}) that
\begin{equation*}\langle\Phi|\overline{F}|\Phi\rangle\leq\sum_{i=1}^n\langle\varphi_i|\varphi_{i}\rangle=\langle\Phi|\Phi\rangle.\end{equation*}
So, $F$ is an operator-valued function.

(2) Secondly, we prove that $F$ is full for the case where all $F_i$ $(1\leq i\leq n)$ are full. It requires us to show that $\overline{F}=  I_{\mathcal{H}_C\otimes\mathcal{H}}$. In fact, for every $1\leq i\leq n$, we have: $$\sum_{\delta_i}F_{i}(\delta_i)^\dag F_{i}(\delta_i)=I_\mathcal{H}$$ because $F_i$ is full. Thus, it follows from equation~(\ref{gud-pr}) that for any $|\Phi\rangle, |\Psi\rangle\in\mathcal{H}_C\otimes\mathcal{H}$,  $$\langle\Phi|\overline{F}|\Psi\rangle=\sum_{i=1}^n\langle\varphi_i|\psi_i\rangle=\langle\Phi|\Psi\rangle.$$ So, it holds that $\overline{F}=I_{\mathcal{H}\otimes\mathcal{H}_s}$ by arbitrariness of $|\Phi\rangle$ and $\Psi\rangle$, and $F$ is full.

\subsection{Proof of Proposition~\ref{qsem}}

Clauses (1) - (4) are obvious. To prove clause (5), let $$P\stackrel{\triangle}{=}\mathbf{measure}\ (\square m\cdot M[\overline{q}:x]=m\rightarrow P_m)\ \mathbf{end}.$$ Then by Definitions \ref{defsem} and \ref{den-def}, for any partial density operator $\rho$ in $\mathcal{H}_{qvar(P)}$, we have: \begin{equation*}\begin{split}\llbracket P\rrbracket (\rho)&=\sum_m\sum_{\delta\in\Delta(P_m)}\lceil P\rceil (\delta[x\leftarrow m])\rho\lceil P\rceil (\delta[x\leftarrow m])^\dag\\
&=\sum_m\sum_{\delta\in\Delta(P_m)}\left(\lceil P_m\rceil (\delta)\otimes I_{qvar(P)\setminus qvar(P_m)}\right)\left(M_m\otimes I_{qvar(P)\setminus\overline{q}}\right)\\ &\ \ \ \ \ \ \ \ \ \ \ \ \ \ \ \ \ \ \ \ \ \ \ \ \ \ \ \ \ \ \ \ \ \rho \left(M_m^\dag\otimes I_{qvar(P)\setminus\overline{q}}\right)
\left(\lceil P_m\rceil (\delta)^\dag\otimes I_{qvar(P)\setminus qvar(P_m)}\right)\\ &=\sum_m\sum_{\delta\in\Delta(P_m)}\left(\lceil P_m\rceil (\delta)\otimes I_{qvar(P)\setminus qvar(P_m)}\right)\left(M_m\rho M_m^\dag\right)\\ &\ \ \ \ \ \ \ \ \ \ \ \ \ \ \ \ \ \ \ \ \ \ \ \ \ \ \ \ \ \ \ \ \ \ \ \ \ \ \ \ \ \ \ \ \ \ \ \ \ \ \ \ \ \ \ \ \ \ \ \ \ \ \ \ \
\left(\lceil P_m\rceil (\delta)^\dag\otimes I_{qvar(P)\setminus qvar(P_m)}\right)\\ &=\sum_m\llbracket P_m\rrbracket \left(M_m\rho M_m^\dag\right)\\ &=\left(\sum_m \left[(M_m\circ M_m^\dag);\llbracket P_m\rrbracket\right]\right)(\rho).
\end{split}\end{equation*}

Finally, we prove clause (6). For simplicity of the presentation, we write: $$P\stackrel{\triangle}{=}\mathbf{qif}\ [\overline{q}](\square i\cdot |i\rangle\rightarrow P_i)\ \mathbf{fiq}.$$ By Definitions \ref{defsem}, we obtain: $$\lceil P\rceil=\square_{i}|i\rangle\rightarrow \lceil P_i\rceil.$$ Note that $\lceil P_i\rceil\in\mathbb{F}(\llbracket P_i\rrbracket)$ for every $1\leq i\leq n$, where $\mathbb{F}(\cdot)$ is defined as in the paragraph before Definition \ref{sfop}. Therefore, it follows from Definition \ref{den-def} that
\begin{equation*}\begin{split}\llbracket P\rrbracket=\mathcal{E}(\lceil P\rceil)\in\{\mathcal{E}(\square_{i}\ |i\rangle\rightarrow F_i):F_i\in\mathbb{F}(\llbracket P_i\rrbracket)\ {\rm for\ every}\ i\}=\square_{i}\ |i\rangle\rightarrow \llbracket P_i\rrbracket.\end{split}
\end{equation*}

\subsection{Proof of Proposition~\ref{wpc}}

The proof is based on the following key lemma by D'Hondt and Panangaden~\cite{DP06}.

\begin{lem}\label{DPL} If the semantic function $\llbracket P\rrbracket$ of program $P$ has the Kraus operator-sum representation: $$\llbracket P\rrbracket =\sum_j E_j\circ E_j^\dag,$$ then we have: $$wp.P=\sum_jE_j^\dag\circ E_j.$$\end{lem}

Now we start to prove Proposition~\ref{wpc}. Clauses (1) - (4) are immediate corollaries of Proposition~\ref{qsem} and Lemma~\ref{DPL}.
 Clause (5) can be directly proved by the transformation between the purely quantum semantics of a program and its weakest precondition semantics given by Lemma \ref{DPL}. We write: $$P\stackrel{\triangle}{=}\mathbf{measure}\ (\square m\cdot M[\overline{q}:x]=m\rightarrow P_m)\ \mathbf{end},$$ and suppose that for every $m$, $$\llbracket P_m\rrbracket =\sum_m E_{mi_m}\circ E_{mi_m}^\dag.$$ Then by Proposition~\ref{qsem} (5) we have: \begin{equation*}\begin{split}
\llbracket P\rrbracket &=\sum_m\left[(M_m\circ M_m^\dag);\llbracket P_m\rrbracket\right]\\
&=\sum_m \left[(M_m\circ M_m^\dag);\sum_{i_m}\left(E_{mi_m}\circ E_{mi_m}^\dag\right)\right]\\
&=\sum_m\sum_{i_m}\left[(E_{mi_m}M_m)\circ (M_m^\dag E_{mi_m}^\dag)\right]\\
&=\sum_m\sum_{i_m}\left[(E_{mi_m}M_m)\circ (E_{mi_m}M_m)^\dag\right].
\end{split}\end{equation*} Using Lemma~\ref{DPL} we obtain:
\begin{equation*}\begin{split}wp.P&=\sum_m\sum_{i_m}\left[(E_{mi_m}M_m)^\dag\circ (E_{mi_m}M_m)\right]\\
&=\sum_m\sum_{i_m}\left[(M_m^\dag E_{mi_m}^\dag)\circ (E_{mi_m}M_m)\right]\\
&=\sum_m\left[\sum_{i_m}\left(E_{mi_m}^\dag\circ E_{mi_m}\right); \left(M_{m}^\dag\circ M_m\right)\right]\\
&=\sum_m \left[wp.P_m;(M_m^\dag\circ M_m)\right].
\end{split}\end{equation*}

 The proof technique of clause (6) is different from that of clause (5). To prove clause (6), it is enough to consider the purely quantum semantics of the involved programs. Instead, we have to go to the semi-classic semantics. For each $1\leq i\leq n$, assume that the semi-classical semantics of $P_i$ is the function $\lceil P_i\rceil$ over $\Delta=\{j_i\}$ such that $$\lceil P_i\rceil (j_i)=E_{ij_i}$$ for every $j_i$. Then
by Definition~\ref{den-def} we obtain:
$$\llbracket P_i\rrbracket =\sum_{j_i}E_{ij_i}\circ E_{ij_i}^\dag,$$ and it follows from Lemma~\ref{DPL} that $$wp.P_i=\sum_{j_i}E_{ij_i}^\dag\circ E_{ij_i}.$$ Now we compute the guarded composition of these operator-valued functions. For any state  $$|\varphi\rangle=\sum_{i=1}^n|i\rangle|\varphi_i\rangle$$ in $\mathcal{H}_C\otimes\mathcal{H}$,  where $|\varphi_i\rangle\in\mathcal{H}_{qvar(P_i)}$ $(1\leq i\leq n)$, we define:
\begin{equation*}\begin{split}G_{j_1...j_n}(|\varphi\rangle)&=\sum_{i=1}^n\zeta_i |i\rangle (E_{ij_i}^\dag |\varphi_i\rangle),\\
\zeta_i&=\prod_{k\neq i}\delta_{kj_k}\end{split}\end{equation*} for any $j_1,...,j_n$ and $i$, where
\begin{equation*}\delta_{kj_k}=\sqrt{\frac{tr (E_{kj_k}^\dag)^\dag E_{kj_k}^\dag}{\sum_{l_k}(E_{kl_k}^\dag)^\dag E_{kl_k}^\dag}}.\end{equation*}
It is obvious that
\begin{equation}\label{coef3}\delta_{kj_k}=\sqrt{\frac{tr E_{kj_k}^\dag E_{kj_k}}{\sum_{l_k} E_{kl_k}^\dag E_{kl_k}}}=\lambda_{kj_k}\end{equation} and $\lambda_{kj_k}$'s are defined by equation~(\ref{coef1}). By Definitions~\ref{qgu} and \ref{def-gsup} we have: $$\sum_{j_1,...,j_n}G_{j_1...j_n}\circ G_{j_1...j_n}^\dag\in\square_{i=1}^n\ |i\rangle\rightarrow wp.P_i.$$ On the other hand, we write $$P\stackrel{\triangle}{=}\mathbf{qif}\ (\square i\cdot |i\rangle\rightarrow P_i)\ \mathbf{fiq}.$$
Then by Definitions \ref{defsem} (5) and~\ref{den-def} we have:
$$\llbracket P\rrbracket=\sum_{j_1,...,j_n}F_{j_1...j_n}\circ F_{j_1...j_n}^\dag$$ where $F_{j_1...j_n}$'s are defined by equation~(\ref{coef0}). Applying Lemma~\ref{DPL} once again, we obtain:
$$wp.P=\sum_{j_1,...,j_n}F_{j_1...j_n}^\dag\circ F_{j_1...j_n}.$$ So, we complete the proof of clause (6) if we are able to prove that $$G_{j_1...j_n}=F_{j_1...j_n}^\dag$$ for all $j_1,...,j_n$. In fact, we can prove the above equality by a straightforward calculation: for any state  $$|\varphi\rangle=\sum_{i=1}^n |i\rangle|\varphi_i\rangle,\ \ \ \ \ \ \ \ \ \ \ \ |\psi\rangle=\sum_{i=1}^n |i\rangle|\psi_i\rangle$$ with $|\varphi\rangle, |\psi_i\rangle\in\mathcal{H}_{qvar(P_i)}$ $(1\leq i\leq n)$, it holds that
\begin{equation*}\begin{split}(G_{j_1...j_n}|\varphi\rangle,|\psi\rangle)&=\left(\sum_{i=1}^n\zeta_i |i\rangle (E_{ij_i}^\dag |\varphi_i\rangle),\sum_{i=1}^n |i\rangle|\psi_i\rangle\right)\\
&=\sum_{i,i^\prime}\zeta_i^\ast\langle i|i^\prime\rangle(E_{ij_i}^\dag |\varphi_i\rangle,|\psi_{i^\prime}\rangle)\\
&=\sum_i\zeta_i(E_{ij_i}^\dag |\varphi_i\rangle,|\psi_{i}\rangle)\\
&=\sum_i\zeta_i(|\varphi_i\rangle,E_{ij_i}|\psi_{i}\rangle)\\
&=\sum_{i,i^\prime}\zeta_i\langle i|i^\prime\rangle(|\varphi_i\rangle,E_{i^\prime j_{i^\prime}}|\psi_{i^\prime}\rangle)\\
&=\left(\sum_{i=1}^n |i\rangle|\varphi_i\rangle,\sum_{i=1}^n \zeta_i |i\rangle (E_{ij_i}|\psi_i\rangle)\right)\\
&=(|\varphi\rangle,F_{j_1...j_n}|\psi\rangle)
\end{split}\end{equation*} because $\zeta_i$'s are real numbers, and it follows from equation~(\ref{coef3}) that $$\zeta_i=\prod_{k\neq i}\lambda_{kj_k}$$ for each $1\leq i\leq n$. Thus, we complete the proof.

\subsection{Proof of Theorem~\ref{proim}}

To simplify the presentation, we write: $$R\stackrel{\triangle}{=}\mathbf{qif}\ [\overline{q}](\square i\cdot |i\rangle\rightarrow P_i)\ \mathbf{fiq}.$$ We need to work at the level of semi-classical semantics first, and then lift it to the purely quantum semantics. Assume that the semi-classical semantics $\lceil P_i\rceil$ is the operator-valued function over $\Delta_i$ such that $\lceil P_i\rceil(\delta_i)=E_{i\delta_i}$ for each $\delta_i\in\Delta_i$. Let states $$|\psi\rangle\in\mathcal{H}_{\bigcup_{i=1}^n qvar(P_i)}$$ and $|\varphi\rangle\in\mathcal{H}_{\overline{q}}$. We can write:
$$|\varphi\rangle=\sum_{i=1}^n\alpha_i |i\rangle$$ for some complex numbers $\alpha_i$ $(1\leq i\leq n)$. Then for any $\delta_i\in\Delta_i$ $(1\leq i\leq n)$, we have: \begin{equation*}\begin{split}
|\Psi_{\delta_1...\delta_n}\rangle&\stackrel{\triangle}{=}\lceil R\rceil (\oplus_{i=1}^n\delta_i)(|\varphi\rangle |\psi\rangle)\\ &=\lceil R\rceil (\oplus_{i=1}^n\delta_i)\left(\sum_{i=1}^n\alpha_i |i\rangle |\psi\rangle\right)\\
&=\sum_{i=1}^n\alpha_i\left(\prod_{k\neq i}\lambda_{k\delta_k}\right)|i\rangle(E_{i\sigma_i}|\psi\rangle)
\end{split}\end{equation*} where $\lambda_{i\delta_i}$'s are defined as in equation~(\ref{coef1}). We continue to compute:
\begin{equation*}\begin{split}
|\Psi_{\delta_1...\delta_n}\rangle\langle\Psi_{\delta_1...\delta_n}|=\sum_{i,j=1}^n \left[\alpha_i\alpha_j^\ast\left(\prod_{k\neq i}\lambda_{k\delta_k}\right)\left(\prod_{k\neq j}\lambda_{k\delta_k}\right) |i\rangle\langle j|\otimes E_{i\delta_i}|\psi\rangle\langle\psi|E_{j\delta_j}^\dag\right],\end{split}\end{equation*} and it follows that
\begin{equation*}\begin{split}
tr_{\mathcal{H}_{\overline{q}}}|\Psi_{\delta_1...\delta_n}\rangle\langle\Psi_{\delta_1...\delta_n}|=\sum_{i=1}^n|\alpha_i|^2 \left(\prod_{k\neq i}\lambda_{k\delta_k}\right)^2 E_{i\delta_i}|\psi\rangle\langle\psi|E_{i\delta_i}^\dag.\end{split}\end{equation*} Using equation~(\ref{coeff}), we obtain:
\begin{equation}\label{trr}\begin{split}tr_{\mathcal{H}_{\overline{q}}}\llbracket R\rrbracket (|\varphi\psi\rangle\langle\psi\varphi|) &=tr_{\mathcal{H}_{\overline{q}}}\left(\sum_{\delta_1,...,\delta_n}|\Psi_{\delta_1...\delta_n}\rangle\langle\Psi_{\delta_1...\delta_n}|\right)\\ &=\sum_{\delta_1,...,\delta_n}tr_{\mathcal{H}_{\overline{q}}}|\Psi_{\delta_1...\delta_n}\rangle\langle\Psi_{\delta_1...\delta_n}|\\
&=\sum_{i=1}^n|\alpha_i|^2\left[\sum_{\delta_1,...,\delta_{i-1},\delta_{i+1},...,\delta_n}\left(\prod_{k\neq i}\lambda_{k\delta_k}\right)^2\right]\cdot \left[\sum_{\delta_i}E_{i\delta_i}|\psi\rangle\langle\psi|E_{i\delta_i}^\dag\right]\\ &=\sum_{i=1}^n|\alpha_i|^2\llbracket P_i\rrbracket (|\psi\rangle\langle\psi|).\end{split}
\end{equation}

Now we do spectral decomposition for $\llbracket P\rrbracket (\rho)$, which is a density operator, and assume that $$\llbracket P\rrbracket (\rho)=\sum_l s_l|\varphi_l\rangle\langle\varphi_l|.$$ We further write: $$|\varphi_l\rangle=\sum_i\alpha_{li}|i\rangle$$ for every $l$. For any density operator $\sigma$ in $\mathcal{H}_{\bigcup_{i=1}^n qvar(P_i)}$, we can write $\sigma$ in the form of $$\sigma=\sum_m r_m|\psi_m\rangle\langle\psi_m|.$$ Then using equation~(\ref{trr}), we get: \begin{equation*}\begin{split}
\llbracket\mathbf{begin\ local}\ \overline{q}:=\rho; &[P]\left(\bigoplus_{i=1}^n |i\rangle\rightarrow P_i\right)\ \mathbf{end}\rrbracket(\sigma)\\ &=tr_{\mathcal{H}_{\overline{q}}}\llbracket P;R\rrbracket (\sigma\otimes\rho)\\
&=tr_{\mathcal{H}_{\overline{q}}}\llbracket R\rrbracket(\sigma\otimes\llbracket P\rrbracket(\rho))\\
&=tr_{\mathcal{H}_{\overline{q}}}\llbracket R\rrbracket\left(\sum_{m,l}r_ms_l|\psi_m\varphi_l\rangle\langle\varphi_l\psi_m|\right)\\
&=\sum_{m,l}r_ms_ltr_{\mathcal{H}_{\overline{q}}}\llbracket R\rrbracket (|\psi_m\varphi_l\rangle\langle\varphi_l\psi_m|)\\
&=\sum_{m,l}r_ms_l\sum_{i=1}^n|\alpha_{li}|^2\llbracket P_i\rrbracket(|\psi_m\rangle\langle\psi_m|)\\
&=\sum_l\sum_{i=1}^ns_l|\alpha_{li}|^2\llbracket P_i\rrbracket\left(\sum_m r_m|\psi_m\rangle\langle\psi_m|\right)\\
&=\sum_l\sum_{i=1}^ns_l|\alpha_{li}|^2\llbracket P_i\rrbracket(\sigma)\\
&=\sum_{i=1}^n\left(\sum_ls_l|\alpha_{li}|^2\right)\llbracket P_i\rrbracket(\sigma)\\
&=\left\llbracket\sum_{i=1}^n P_i@p_i\right\rrbracket(\sigma),
\end{split}\end{equation*} where \begin{equation*}\begin{split}
p_i=\sum_{l}s_l|\alpha_{li}|^2
=\sum_ls_l\langle i|\varphi_l\rangle\langle\varphi_l|i\rangle=\langle i|\left(\sum_ls_l|\varphi_l\rangle\langle\varphi_l|\right)|i\rangle=\langle i|\llbracket P\rrbracket (\rho)|i\rangle.\end{split}\end{equation*}

\subsection{Proof of Theorem~\ref{local}}

We first prove equation~(\ref{local1}). Let $LHS$ and $RHS$ stand for the left and right hand side of equation~(\ref{local1}), respectively. What we want to prove is $\llbracket LHS\rrbracket =\llbracket RHS\rrbracket$. But we need to work with the semi-classical semantics, and show that $\lceil LHS\rceil=\lceil RHS\rceil$. Assume that $\lceil P_i\rceil$ is the operator-valued function over $\Delta_i$ such that $$\lceil P_i\rceil (\delta_i)=F_{i\delta_i}$$ for each $\delta_i\in\Delta_i$ $(1\leq i\leq n)$. We write: $$P\stackrel{\triangle}{=}\mathbf{qif}\ (\square i\cdot  U_{\overline{q}}^\dag |i\rangle\rightarrow P_i)\ \mathbf{fiq}.$$ Then for any state $$|\psi\rangle=\sum_{i=1}^n |i\rangle|\psi_i\rangle,$$ where $|\psi_i\rangle\in\mathcal{H}_V$ $(1\leq i\leq n)$, and $V=\bigcup_{i=1}^n qvar(P_i)$, we have: \begin{equation*}\begin{split}
\lceil P\rceil(\oplus_{i=1}^n\delta_i)|\psi\rangle &=\lceil P\rceil(\oplus_{i=1}^n\delta_i)\left[\sum_{i=1}^n\left(\sum_{j=1}^n U_{ij}(U_{\overline{q}}^\dag |j\rangle)\right)|\psi_i\rangle\right]
\\ &=\lceil P\rceil(\oplus_{i=1}^n\delta_i)\left[\sum_{j=1}^n(U_{\overline{q}}^\dag |j\rangle)\left(\sum_{i=1}^n U_{ij}|\psi_i\rangle\right)\right]
\\ &=\sum_{j=1}^n \left(\prod_{k\neq j}\lambda_{k\delta_k}\right)(U_{\overline{q}}^\dag |j\rangle) F_{j\delta_j}\left(\sum_{i=1}^n U_{ij}|\psi_i\rangle\right),
\end{split}\end{equation*} where $\lambda_{k\delta_k}$'s are defined by equation (\ref{coef1}). Then it holds that \begin{equation*}\begin{split}
\lceil RHS\rceil (\oplus_{i=1}^n\delta_i)|\psi\rangle &= U_{\overline{q}}(\lceil P\rceil (\oplus_{i=1}^n\delta_i)|\psi\rangle)\\
& =\sum_{j=1}^n \left(\prod_{k\neq j}\lambda_{k\delta_k}\right)|j\rangle F_{j\delta_j}\left(\sum_{i=1}^n U_{ij}|\psi_i\rangle\right)\\
&=\lceil P\rceil (\oplus_{i=1}^n\delta_i)\left[\sum_{j=1}^n|j\rangle\left(\sum_{i=1}^n U_{ij}|\psi_i\rangle\right)\right]\\
&=\lceil P\rceil (\oplus_{i=1}^n\delta_i)\left[\sum_{i=1}^n \left(\sum_{j=1}^nU_{ij}|j\rangle\right)|\psi_i\rangle\right]\\
&=\lceil P\rceil (\oplus_{i=1}^n\delta_i)\left(\sum_{i=1}^n (U_{\overline{q}}|i\rangle)|\psi_i\rangle\right)\\
&=\lceil LHS\rceil (\oplus_{i=1}^n\delta_i) |\psi\rangle.
\end{split}\end{equation*} So, we complete the proof of equation~(\ref{local1}).

Now we are ready to prove equation~(\ref{local2}). The basic idea is to use equation~(\ref{local1}) that we just proved above to prove the more general equation~(\ref{local2}). So, we need to turn the general \textquotedblleft coin\textquotedblright\ program $P$ into a special \textquotedblleft coin\textquotedblright\ program which is a unitary transformation. The technique that we used before to deal with super-operators is always the Kraus operator-sum representation. Here, however, we have to employ the system-environment model of super-operators (see equation~(8.38) in \cite{NC00}). Since $\llbracket P\rrbracket$ is a super-operator in $\mathcal{H}_{\overline{q}}$, there must be a family of quantum variables $\overline{r}$, a pure state $|\varphi_0\rangle\in\mathcal{H}_{\overline{r}}$, a unitary operator $U$ in $\mathcal{H}_{\overline{q}}\otimes\mathcal{H}_{\overline{r}}$, and a projection operator $K$ onto some closed subspace $\mathcal{K}$ of $\mathcal{H}_{\overline{r}}$ such that \begin{equation}\label{localx}\llbracket P\rrbracket (\rho)=tr_{\mathcal{H}_{\overline{r}}}(KU(\rho\otimes |\varphi_0\rangle\langle\varphi_0|)U^\dag K)\end{equation} for all density operators $\rho$ in $\mathcal{H}_{\overline{q}}$. We choose an orthonormal basis of $\mathcal{K}$ and then extend it to an orthonormal basis $\{|j\rangle\}$ of $\mathcal{H}_{\overline{r}}$. Define pure states $|\psi_{ij}\rangle=U^\dag |ij\rangle$ for all $i, j$ and programs $$Q_{ij}=\begin{cases}P_i\ &{\rm if}\ |j\rangle\in\mathcal{K},\\ \mathbf{abort}\ &{\rm if}\ |j\rangle\notin \mathcal{K}.\end{cases}$$ Then by a routine calculation we have:
\begin{equation}\label{local3}\llbracket\mathbf{qif}\ (\square i,j\cdot |ij\rangle\rightarrow Q_{ij})\ \mathbf{fiq}\rrbracket (\sigma)=\llbracket \mathbf{qif}\ (\square i\cdot |i\rangle\rightarrow P_i)\ \mathbf{fiq}\rrbracket (K\sigma K)
\end{equation} for any $\sigma\in\mathcal{H}_{\overline{q}\cup\overline{r}\cup V}$, where $V=\bigcup_{i=1}^n qvar(P_i)$. We now write $RHS$ for the right hand side of equation~(\ref{local2}). Then we have: \begin{equation*}\begin{split}\llbracket RHS\rrbracket (\rho)&=tr_{\mathcal{H}_{\overline{r}}}\left(\llbracket\mathbf{qif}\ (\square i,j\cdot U^\dag |ij\rangle\rightarrow Q_{ij})\ \mathbf{fiq};U[\overline{q},\overline{r}]\rrbracket (\rho\otimes |\varphi_0\rangle\langle\varphi_0|\right)\\ & =tr_{\mathcal{H}_{\overline{r}}}\left(\left\llbracket [U[\overline{q}, \overline{r}]]\left(\bigoplus_{i,j} |ij\rangle\rightarrow Q_{ij}\right) \right\rrbracket (\rho\otimes |\varphi_0\rangle\langle\varphi_0|)\right)\\ & =tr_{\mathcal{H}_{\overline{r}}}\left(\llbracket\mathbf{qif}\ (\square i,j\cdot |ij\rangle\rightarrow Q_{ij})\ \mathbf{fiq}\rrbracket (U(\rho\otimes |\varphi_0\rangle\langle\varphi_0|)U^\dag)\right)\\
& =tr_{\mathcal{H}_{\overline{r}}}\llbracket\mathbf{qif}\ (\square i\cdot |i\rangle\rightarrow P_{i})\ \mathbf{fiq}\rrbracket (KU(\rho\otimes |\varphi_0\rangle\langle\varphi_0|)U^\dag K)\\
& =\llbracket\mathbf{qif}\ (\square i\cdot |i\rangle\rightarrow P_{i})\ \mathbf{fiq}\rrbracket (tr_{\mathcal{H}_{\overline{r}}}(KU(\rho\otimes |\varphi_0\rangle\langle\varphi_0|)U^\dag K))\\
& =\llbracket \mathbf{qif}\ (\square i\cdot |i\rangle\rightarrow P_{i})\ \mathbf{fiq}\rrbracket (\llbracket P\rrbracket (\rho))\\ & =\left\llbracket [P]\left(\bigoplus_{i} |i\rangle\rightarrow P_{i}\right)\right\rrbracket (\rho)
\end{split}\end{equation*} for all density operators $\rho$ in $\mathcal{H}_{\overline{q}}$.
Here, the second equality is obtained by using equation~(\ref{local1}), the fourth equality comes from (\ref{local3}), the fifth equality holds because $\overline{r}\cap qvar(\mathbf{qif}\ (\square i\cdot |i\rangle\rightarrow P_{i})\ \mathbf{fiq})=\emptyset$, and the sixth equality follows from equation (\ref{localx}). Therefore, equation~(\ref{local2}) is proved.

\subsection{Proof of Theorem~\ref{law0} and Theorem~\ref{law}}

The proof of Theorem~\ref{law0} is similar to but simpler than the proof of Theorem \ref{law}. So, here we only prove Theorem \ref{law}.

(1) Clause (1) is immediate from Theorem~\ref{proim}.

(2) To prove clause (2), we write: \begin{equation*}\begin{split}Q&\stackrel{\triangle}{=}\mathbf{qif}\ [\overline{q}](\square i\cdot |i\rangle\rightarrow P_i)\ \mathbf{fiq},\\
R&\stackrel{\triangle}{=}\mathbf{qif}\ [\overline{q}](\square i\cdot |i\rangle\rightarrow P_{\tau(i)})\ \mathbf{fiq}.
\end{split}\end{equation*} By definition, we have $LHS=P;R$ and $RHS=P;U_\tau[\overline{q}];Q;U_{\tau^{-1}}[\overline{q}].$ So, it suffices to show that $R\equiv U_\tau[\overline{q}];Q;U_{\tau^{-1}}[\overline{q}].$ Again, we first need to deal with the semi-classical semantics of the two sides of this equality. Assume that $\lceil P_i\rceil$ is the operator-valued function over $\Delta_i$ with $\lceil P_i\rceil(\delta_i)=E_{i\delta_i}$ for each $\delta_i\in\Delta_i$ $(1\leq i\leq n)$. For each state $|\Psi\rangle\in\mathcal{H}_{\overline{q}\cup\bigcup_{i=1}^nqvar(P_i)}$, we can write: $$|\Psi\rangle=\sum_{i=1}^n |i\rangle|\psi_i\rangle$$ for some $|\psi_i\rangle\in\mathcal{H}_{\bigcup_{i=1}^nqvar(P_i)}$ $(1\leq i\leq n)$. Then for any $\delta_1\in\Delta_{\tau(1)},...,\delta_n\in\Delta_{\tau(n)}$, it holds that
\begin{equation*}\begin{split}|\Psi_{\delta_1...\delta_n}\rangle &\stackrel{\triangle}{=}\lceil R\rceil(\oplus_{i=1}^n\delta_i)(|\Psi\rangle)\\ &=\sum_{i=1}^n\left(\prod_{k\neq i}\mu_{k\delta_k}\right)|i\rangle (E_{\tau(i)\delta_i}|\psi_i\rangle),\end{split}\end{equation*} where \begin{equation}\label{ceoceo}\mu_{k\delta_k}=\sqrt{\frac{tr E^\dag_{\tau(k)\delta_k}E_{\tau(k)\delta_k}}{\sum_{\theta_k\in\Sigma_{\tau(k)}}tr E^\dag_{\tau(k)\theta_k}E_{\tau(k)\theta_k}}}=\lambda_{\tau(k)\delta_k}\end{equation} for every $k$ and $\delta_k$, and $\lambda_{i\sigma_i}$'s are defined by equation~(\ref{coef1}). On the other hand, we first observe: $$|\Psi^\prime\rangle\stackrel{\triangle}{=}(U_\tau)_{\overline{q}}(|\Psi\rangle)=\sum_{i=1}^n|\psi_i\rangle|\tau(i)\rangle=\sum_{j=1}^n|\psi_{\tau^{-1}(j)}\rangle|j\rangle.$$ Then for any $\delta_1\in\Delta_1,...,\delta_n\in\Delta_n$, it holds that \begin{equation*}\begin{split}|\Psi_{\delta_1...\delta_n}^{\prime\prime}\rangle&\stackrel{\triangle}{=}\lceil Q\rceil \left(\bigoplus_{i=1}^n\delta_i\right)\left(|\Psi^\prime\rangle\right)\\ &=\sum_{j=1}^n\left(\prod_{l\neq j}\lambda_{l\delta_{\tau^{-1}(l)}}\right)|j\rangle(E_{j\delta_{\tau^{-1}(j)}}|\psi_{\tau^{-1}(j)}\rangle)\\ &=\sum_{i=1}^n\left(\prod_{k\neq i}\lambda_{\tau(k)\delta_{k}}\right)|\tau(i)\rangle(E_{\tau(i)\delta_i}|\psi_{i}\rangle).\end{split}\end{equation*} Furthermore, we have: $$(U_{\tau^{-1}})_{\overline{q}}(|\Psi_{\delta_1...\delta_n}^{\prime\prime}\rangle)=\sum_{i=1}^n\left(\prod_{k\neq i}\lambda_{\tau(k)\delta_{k}}\right)|i\rangle(E_{\tau(i)\delta_i}|\psi_{i}\rangle).$$ Therefore, we can compute the purely quantum semantics:  \begin{equation}\label{com-den}\begin{split}\llbracket U_\tau[\overline{q}];Q;U_{\tau^{-1}}[\overline{q}]\rrbracket(|\Psi\rangle\langle\Psi|)&=\llbracket Q;U_{\tau^{-1}}[\overline{q}](|\Psi^\prime\rangle\langle\Psi^\prime|)\rrbracket\\ &=(U_{\tau^{-1}})_{\overline{q}}\left(\sum_{\delta_1,...,\delta_n}(|\Psi^{\prime\prime}_{\delta_1...\delta_n}\rangle\langle\Psi^{\prime\prime}_{\delta_1...\delta_n}|\right)(U_\tau)_{\overline{q}}\\ &=\sum_{\delta_1,...,\delta_n}|\Psi_{\delta_1...\delta_n}\rangle\langle\Psi_{\delta_1...\delta_n}|\\ &=\llbracket R\rrbracket (\Psi\rangle\langle\Psi|).
\end{split}\end{equation} Here, the second equality comes from equation (\ref{ceoceo}) and the fact that $\tau$ is one-onto-one, and thus $\tau^{-1}(j)$ traverses over $1,...,n$ as $j$ does. Thus, it follows from equation~(\ref{com-den}) and spectral decomposition that $$\llbracket R\rrbracket (\rho)=\llbracket U_\tau[\overline{q}];Q;U_{\tau^{-1}}[\overline{q}]\rrbracket(\rho)$$ for any density operator $\rho$ in $\mathcal{H}_{\overline{q}\cup\bigcup_{i=1}^nqvar(P_i)}$, and we complete the proof of clause (2).

(3) To prove clause (3), we write: \begin{equation*}\begin{split}X_i&\stackrel{\triangle}{=}\mathbf{qif}\ (\square j_i\cdot |j_i\rangle\rightarrow R_{ij_i})\ \mathbf{fiq},\\
Y_i&\stackrel{\triangle}{=}[Q_i]\left(\bigoplus_{j_i=1}^{n_i}|j_i\rangle\rightarrow R_{ij_i}\right)
\end{split}\end{equation*} for every $1\leq i\leq m$, and we further put:
\begin{equation*}\begin{split}X&\stackrel{\triangle}{=}\mathbf{qif}\ (\square i\cdot |i\rangle\rightarrow Y_i)\ \mathbf{fiq},\\
T&\stackrel{\triangle}{=}\mathbf{qif}\ (\square i\cdot |i\rangle\rightarrow Q_i)\ \mathbf{fiq},\\
Z&\stackrel{\triangle}{=}\mathbf{qif}\ (\overline{\alpha})(\square i,j_i\in\Delta\cdot |i,j_i\rangle\rightarrow R_{ij_i})\ \mathbf{fiq}.\\
\end{split}\end{equation*} Then by the definition of quantum choice we have $LHS=P;X$ and $RHS=P;T;Z$. So, it suffices to show that $X\equiv T;Z$. To do this, we consider the semi-classical semantics of the involved programs. For each $1\leq i\leq m$, and for each $1\leq j_i\leq n_i$, we assume: \begin{itemize}\item $\lceil Q_i\rceil$ is the operator-valued function over $\Delta_i$ such that $\lceil Q_i\rceil(\delta_i)=F_{i\delta_i}$ for every $\delta_i\in\Delta_i$; and \item $\lceil R_{ij_i}\rceil$ is the operator-valued function over $\Sigma_{ij_i}$ such that $\lceil R_{ij_i}\rceil(\sigma_{ij_i})=E_{(ij_i)\sigma_{ij_i}}$ for every $\sigma_{ij_i}\in\Sigma_{ij_i}$.\end{itemize} We also assume that state  $$|\Psi\rangle=\sum_{i=1}^m |i\rangle|\Psi_i\rangle$$ where each $|\Psi_i\rangle$ is further decomposed into $$|\Psi_i\rangle=\sum_{j_i=1}^{n_i}|j_i\rangle|\psi_{ij_i}\rangle$$ with $|\psi_{ij_i}\rangle\in\mathcal{H}_{\bigcup_{j_i=1}^{n_i}qvar(R_{ij_i})}$ for every $1\leq i\leq m$ and $1\leq j_i\leq n_i$. To simplify the presentation, we use the abbreviation  $\overline{\sigma}_i=\oplus_{j_i=1}^{n_i}\sigma_{ij_i}$. Now we compute the semi-classical semantics of program $Y_i$: \begin{equation}\label{use1}\begin{split}\lceil Y_i\rceil(\delta_i\overline{\sigma}_i)|\Psi_i\rangle&=\lceil X_i\rceil(\overline{\sigma}_i)(\lceil Q_i\rceil(\delta_i)|\Psi_i\rangle)\\
&=\lceil X_i\rceil(\overline{\sigma}_i)\left(\sum_{j_i=1}^{n_i}\left(F_{i\delta_i}|j_i\rangle\right)|\psi_{ij_i}\rangle \right)\\ &=\lceil X_i\rceil(\overline{\sigma}_i)\left[\sum_{j_i=1}^{n_i}\left(\sum_{l_i=1}^{n_i}\langle l_i|F_{i\delta_i}|j_i\rangle|l_i\rangle\right)|\psi_{ij_i}\rangle \right]\\ &=\lceil X_i\rceil(\overline{\sigma}_i)\left[\sum_{l_i=1}^{n_i}|l_i\rangle\left(\sum_{j_i=1}^{n_i}\langle l_i|F_{i\delta_i}|j_i\rangle|\psi_{ij_i}\rangle\right)\right]\\ &=\sum_{l_i=1}^{n_i}\left[
\Lambda_{il_i}\cdot|l_i\rangle\left(\sum_{j_i=1}^{n_i}\langle l_i|F_{i\delta_i}|j_i\rangle E_{(il_i)\sigma_{il_i}}|\psi_{ij_i}\rangle\right)\right]
\end{split}\end{equation} where the coefficients:
$$\Lambda_{il_i}=\prod_{l\neq l_i}\lambda_{(il)\sigma_{il}},$$ $$\lambda_{(il)\sigma_{il}}=\sqrt{\frac{tr E^\dag_{(il)\sigma_{il}}E_{(il)\sigma_{il}}}{\sum_{k=1}^{n_i}tr E^\dag_{(ik)\sigma_{ik}}E_{(ik)\sigma_{ik}}}}$$ for each $1\leq l\leq n_i$. Then using equation (\ref{use1}), we can further compute the semi-classical semantics of program $X$:
\begin{equation}\label{use2}\begin{split}\lceil X\rceil (\oplus_{i=1}^m(\delta_i\overline{\sigma}_i))|\Psi\rangle&=
\sum_{i=1}^m \left(\Gamma_i \cdot|i\rangle \lceil Y_i\rceil (\delta_i\overline{\sigma}_i)|\Psi_i\rangle\right)\\ &=\sum_{i=1}^m\sum_{l_i=1}^{n_i}\left[\Gamma_i\cdot\Lambda_{il_i}\cdot|il_i\rangle\left(\sum_{j_i=1}^{n_i}\langle l_i|F_{i\delta_i}|j_i\rangle E_{(il_i)\sigma_{il_i}}|\psi_{ij_i}\rangle\right)\right]
\end{split}\end{equation}
where $$\Gamma_i=\prod_{h\neq i}\gamma_{h\overline{\sigma}_h},$$ \begin{equation}\label{ggmm}\gamma_{i\overline{\sigma}_i}=\sqrt{\frac{tr\lceil Y_i\rceil(\delta_i\overline{\sigma}_i)^\dag \lceil Y_i\rceil(\delta_i\overline{\sigma}_i)}{\sum_{h=1}^mtr\lceil Y_h\rceil(\delta_h\overline{\sigma}_h)^\dag \lceil Y_h\rceil (\delta_h\overline{\sigma}_h)}}.\end{equation}

On the other hand, we can compute the semi-classical semantics of program $T$: \begin{equation*}\begin{split}\lceil T\rceil(\oplus_{i=1}^m \delta_i)|\Psi\rangle &=\lceil T\rceil (\oplus_{i=1}^m\delta_i)\left(\sum_{i=1}^m |i\rangle|\Psi_i\rangle\right)\\ &=
\sum_{i=1}^m\left(\Theta_i\cdot |i\rangle F_{i\delta_i}|\Psi_i\rangle\right)\\ &=\sum_{i=1}^m\left[\Theta_i\cdot |i\rangle\left(\sum_{j_i=1}^{n_i}(F_{i\delta_i}|j_i\rangle)|\psi_{ij_i}\rangle\right)\right]\\
&=\sum_{i=1}^m\left[\Theta_i\cdot |i\rangle\left(\sum_{j_i=1}^{n_i}\left(\sum_{l_i=1}^{n_i}\langle l_i|F_{i\delta_i}|j_i\rangle|l_i\rangle\right)|\psi_{ij_i}\rangle\right)\right]\\
&=\sum_{i=1}^m \sum_{l_i=1}^{n_i}\left[\Theta_i\cdot |il_i\rangle\left(\sum_{j_i=1}^{n_i}\langle l_i|F_{i\delta_i}|j_i\rangle|\psi_{ij_i}\rangle\right)\right]
\end{split}\end{equation*} where $$\Theta_i=\prod_{h\neq i}\theta_{h\delta_h},$$ $$\theta_{i\delta_i}=\sqrt{\frac{tr F_{i\delta_i}^\dag E_{i\delta_i}}{\sum_{h=1}^m tr F_{h\delta_h}^\dag F_{h\delta_h}}}$$ for every $1\leq i\leq m$.
Consequently, we obtain the semi-classical semantics of program $T;Z$: \begin{equation}\label{use3}\begin{split}\lceil T;Z\rceil (&(\oplus_{i=1}^m\delta_i)(\oplus_{i=1}^m\overline{\sigma}_{i}))|\Psi\rangle\\
&=\lceil Z\rceil (\oplus_{i=1}^m\overline{\sigma}_{i})(\lceil T\rceil (\oplus_{i=1}^m\delta_i)|\Psi\rangle)\\
&=\lceil Z\rceil (\oplus_{i=1}^m\oplus_{l_i=1}^{n_i}\sigma_{ij_i})\\ &\ \ \ \ \ \ \ \ \ \ \left(\sum_{i=1}^m\sum_{l_i=1}^{n_i}\left[\Theta_i\cdot |il_i\rangle \left(\sum_{j_i=1}^{n_i}\langle l_i|F_{i\delta_i}|j_i\rangle|\psi_{ij_i}\rangle\right)\right]\right)\\ &=\sum_{i=1}^m\sum_{l_i=1}^{n_i}\left[\alpha^{il_i}_{\{\sigma_{jk_j}\}_{(j,k_j)\neq (i,l_i)}}\cdot\Theta_i\cdot |il_i\rangle \left(\sum_{j_i=1}^{n_i}\langle l_i|F_{i\delta_i}|j_i\rangle E_{(il_i)\sigma_{il_i}}|\psi_{ij_i}\rangle\right)\right].
\end{split}\end{equation} By comparing equations (\ref{use2}) and (\ref{use3}), we see that it suffices to take \begin{equation}\label{use4}\alpha^{il_i}_{\{\sigma_{jk_j}\}_{(j,k_j)\neq (i,l_i)}}=\frac{\Gamma_i\cdot\Delta_{il_i}}{\Theta_i}\end{equation} for all $i, l_i$ and $\{\sigma_{jk_j}\}_{(j,k_j)\neq (i,l_i)}$. What remains to prove is the normalization condition: \begin{equation}\label{use5}
\sum_{\{\sigma_{jk_j}\}_{(j,k_j)\neq (i,l_i)}} \left|\alpha^{il_i}_{\{\sigma_{jk_j}\}_{(j,k_j)\neq (i,l_i)}}\right|^2=1.
\end{equation} To do this, we first compute coefficients $\gamma_{i\overline{\sigma}_i}$. Let $\{|\varphi\rangle\}$ be an orthonormal basis of $\mathcal{H}_{\bigcup_{j_i=1}^{n_i}qvar(R_{ij_i})}$. Then we have:
$$G_{\varphi j_i}\stackrel{\triangle}{=}\lceil Y_i\rceil(\delta_i\overline{\sigma}_i)|\varphi\rangle|j_i\rangle=\sum_{l_i=1}^{n_i}\Lambda_{il_i}\cdot \langle l_i|F_{i\delta_i}|j_i\rangle E_{(il_i)\sigma_{il_i}}|\varphi\rangle|l_i\rangle.$$
It follows that \begin{equation*}\begin{split}G_{\varphi j_i}^\dag G_{\varphi j_i}&=\sum_{l_i,l^\prime_i=1}^{n_i}\Lambda_{il_i}\cdot \Lambda_{il^\prime_i}
\langle j_i|F^\dag_{i\delta_i}|l_i\rangle
\langle l^\prime_i|F_{i\delta_i}|j_i\rangle \langle\varphi|E_{(il_i)\sigma_{il_i}}^\dag E_{(il^\prime_i)\sigma_{il^\prime_i}}|\varphi\rangle\langle l_i|l^\prime_i\rangle\\
&=\sum_{l_i=1}^{n_i}\Lambda_{il_i}^2\cdot
\langle j_i|F^\dag_{i\delta_i}|l_i\rangle
\langle l_i|F_{i\delta_i}|j_i\rangle \langle\varphi|E_{(il_i)\sigma_{il_i}}^\dag E_{(il_i)\sigma_{il_i}}|\varphi\rangle.
\end{split}\end{equation*} Furthermore, we obtain:
\begin{equation}\label{use6}\begin{split}
tr\lceil Y_i\rceil&(\delta_i\overline{\sigma}_i)^\dag\lceil Y_i\rceil(\delta_i\overline{\sigma}_i)=\sum_{\varphi, j_i}G_{\varphi j_i}^\dag G_{\varphi j_i}\\
&=
\sum_{l_i=1}^{n_i}\Lambda_{il_i}^2\cdot
\left(\sum_{j_i}\langle j_i|F^\dag_{i\delta_i}|l_i\rangle
\langle l_i|F_{i\delta_i}|j_i\rangle\right)\left(\sum_{\varphi}\langle\varphi|E_{(il_i)\sigma_{il_i}}^\dag E_{(il_i)\sigma_{il_i}}|\varphi\rangle\right)\\
&=
\sum_{l_i=1}^{n_i}\Lambda_{il_i}^2\cdot
tr (F^\dag_{i\delta_i}|l_i\rangle
\langle l_i|F_{i\delta_i}) tr (E_{(il_i)\sigma_{il_i}}^\dag E_{(il_i)\sigma_{il_i}}).
\end{split}\end{equation} Now a routine but tedious calculation yields equation (\ref{use5}) through substituting equation (\ref{use6}) into (\ref{ggmm}) and then substituting equations (\ref{ggmm}) and (\ref{use4}) into (\ref{use5}).

(4) Finally, we prove clause (4). To prove the first equality, we write: \begin{equation*}\begin{split}
X&\stackrel{\triangle}{=}\mathbf{qif}\ (\square i\cdot |i\rangle\rightarrow P_i)\ \mathbf{fiq},\\ Y&\stackrel{\triangle}{=}\mathbf{qif}\ (\overline{\alpha})(\square i\cdot |i\rangle\rightarrow (P_i;Q))\ \mathbf{fiq}.
\end{split}\end{equation*} Then by definition we have $LHS=P;X;Q$ and $RHS=P;Y$. So, it suffices to show that $X;Q\equiv_{CF}Y$. Suppose that $\lceil P_i\rceil (\sigma_i)=E_{i\sigma_i}$ for every $\sigma_i\in\Delta(P_i)$ and $\lceil Q\rceil (\delta)=F_\delta$ for every $\delta\in\Delta(Q),$ and suppose that $$|\Psi\rangle=\sum_{i=1}^n|i\rangle|\psi_i\rangle,$$ where $|\psi_i\rangle\in\mathcal{H}_{\bigcup_i qvar(P_i)}$ for all $i$. Then it holds that \begin{equation*}\begin{split}
\lceil X;Q\rceil ((\oplus_{i=1}^n \sigma_i)\delta)|\Psi\rangle&=\lceil Q\rceil(\delta)(\lceil X\rceil(\oplus_{i=1}^n\sigma_i)|\Psi\rangle)\\ &=F_\delta\left(\sum_{i=1}^n\Lambda_i |i\rangle(E_{i\sigma_i}|\psi_i\rangle)\right)\\ &=\sum_{i=1}^n\Lambda_i\cdot |i\rangle(F_\delta E_{i\sigma_i}|\psi_i\rangle)
\end{split}\end{equation*} because $qvar(P)\cap qvar(Q)=\emptyset$, where $$\Lambda_i=\prod_{k\neq i}\lambda_{k\sigma_k},$$ \begin{equation}\label{cceeoo}\lambda_{i\sigma_i}=\sqrt{\frac{tr E_{i\sigma_i}^\dag E_{i\sigma_i}}{\sum_{k=1}^n tr E_{k\sigma_k}^\dag E_{k\sigma_k}}}.\end{equation}
Furthermore, we have: \begin{equation}\label{tttra}\begin{split}tr_{\mathcal{H}_{qvar(P)}}&(\llbracket X;Q\rrbracket (|\Psi\rangle\langle\Psi|))\\ &=
tr_{\mathcal{H}_{qvar(P)}}\left[\sum_{\{\sigma_i\},\delta}\sum_{i,j}\Lambda_i\Lambda_j\cdot |i\rangle\langle j|(F_\delta E_{i\sigma_i}|\psi_i\rangle\langle\psi_j|E_{j\sigma_j}^\dag F_\delta^\dag)\right]\\
&=\sum_{\{\sigma_i\},\delta}\sum_{i}\Lambda_i^2\cdot F_\delta E_{i\sigma_i}|\psi_i\rangle\langle\psi_i|E_{i\sigma_i}^\dag F_\delta^\dag.\end{split}\end{equation}
On the other hand, we can compute the semi-classical semantics of $Y$:
\begin{equation*}\begin{split}\lceil Y\rceil (\oplus_{i=1}^n\sigma_i\delta_i)|\Psi\rangle&=\sum_{i=1}^n\alpha^{(i)}_{\{\sigma_k,\delta_k\}_{k\neq i}}\cdot|i\rangle(\lceil P_i;Q\rceil(\sigma_i\delta_i)|\psi_i\rangle)\\
&=\sum_{i=1}^n\alpha^{(i)}_{\{\sigma_k,\delta_k\}_{k\neq i}}\cdot|i\rangle(F_{\delta_i} E_{\sigma_i}|\psi_i\rangle).\end{split}\end{equation*} Furthermore, we obtain: \begin{equation}\label{tttra1}\begin{split}&tr_{\mathcal{H}_{qvar(P)}}(\llbracket Y\rrbracket (|\Psi\rangle\langle\Psi|))\\ &=
tr_{\mathcal{H}_{qvar(P)}}\left[\sum_{\{\sigma_i,\delta_i\}}\sum_{i,j}\alpha^{(i)}_{\{\sigma_k,\delta_k\}_{k\neq i}}(\alpha^{(j)}_{\{\sigma_l,\delta_l\}_{l\neq j}})^\ast\cdot |i\rangle\langle j|(F_{\delta_i} E_{i\sigma_i}|\psi_i\rangle\langle\psi_j|E_{j\sigma_j}^\dag F_{\delta_j}^\dag)\right]\\
&=\sum_{\{\sigma_i\},\delta}\sum_{i}\left|\alpha^{(i)}_{\{\sigma_k,\delta_k\}_{k\neq i}}\right|^2\cdot F_{\delta_i} E_{i\sigma_i}|\psi_i\rangle\langle\psi_i|E_{i\sigma_i}^\dag F_{\delta_i}^\dag.\end{split}\end{equation} Comparing equations (\ref{tttra}) and (\ref{tttra1}), we see that $$tr_{\mathcal{H}_{qvar(P)}}(\llbracket X;Q\rrbracket (|\Psi\rangle\langle\Psi|))=tr_{\mathcal{H}_{qvar(P)}}(\llbracket Y\rrbracket (|\Psi\rangle\langle\Psi|))$$ if we take $$\alpha^{(i)}_{\{\sigma_k,\delta_k\}_{k\neq i}}=\frac{\Lambda_i}{\sqrt{|\Delta(Q)|}}$$ for all $i, \{\sigma_k\}$ and $\{\delta_k\}$. Since $qvar(P)\subseteq cvar(X;Q)\cup cvar(Y)$, it follows that $$tr_{\mathcal{H}_{cvar(X;Q)\cup cvar(Y)}}(\llbracket X;Q\rrbracket (|\Psi\rangle\langle\Psi|))=tr_{\mathcal{H}_{cvar(X;Q)\cup cvar(Y)}}(\llbracket Y\rrbracket (|\Psi\rangle\langle\Psi|)).$$ Therefore, we can assert that $$tr_{\mathcal{H}_{cvar(X;Q)\cup cvar(Y)}}(\llbracket X;Q\rrbracket (\rho))=tr_{\mathcal{H}_{cvar(X;Q)\cup cvar(Y)}}(\llbracket Y\rrbracket (\rho))$$ for all density operator $\rho$ by spectral decomposition, and $X;Q\equiv_{CF} Y$.

For the special case where $Q$ contains no measurements, $\Delta(Q)$ is a singleton, say $\{\delta\}$. We write: $$Z\stackrel{\triangle}{=}\mathbf{qif}\ (\square i\cdot |i\rangle\rightarrow (P_i;Q))\ \mathbf{fiq}.$$ Then
\begin{equation*}\lceil Z\rceil (\oplus_{i=1}^n\sigma_i\delta)|\Psi\rangle =\sum_{i=1}^n \left(\prod_{k\neq i}\theta_{k\sigma_k}\right)\cdot|i\rangle(F_{\delta} E_{\sigma_i}|\psi_i\rangle),\end{equation*} where \begin{equation*}\theta_{i\sigma_i}=\sqrt{\frac{tr E_{i\sigma_i}^\dag F_\delta^\dag F_\delta E_{i\sigma_i}}{\sum_{k=1}^n tr E_{k\sigma_k}^\dag F_\delta^\dag F_\delta E_{k\sigma_k}}}=\lambda_{i\sigma_i},\end{equation*} where $\lambda_{i\sigma_i}$ is given by equation (\ref{cceeoo}), because $F_\delta^\dag F_\delta$ is the identity operator. Consequently, $\lceil X;Q\rceil =\lceil Z\rceil$, and we complete the proof of the second equality of clause (4).

\end{document}